\documentclass[reprint,nofootinbib,superscriptaddress,amsmath,amssymb,aps,prd]{revtex4-1}
\usepackage{amsmath,mathrsfs,amsbsy,color,graphicx,bm,amsthm,amsfonts}
\usepackage{bbm}
\usepackage{times}
\usepackage{units}
\usepackage{dcolumn}
\usepackage{graphicx}
\usepackage{epsfig}
\usepackage{epstopdf}
\usepackage[colorlinks,linkcolor=red,anchorcolor=green,citecolor=blue,CJKbookmarks=True]{hyperref}
\DeclareMathSymbol{\shortminus}{\mathbin}{AMSa}{"39}
\usepackage{mathrsfs}
\usepackage{braket}
\usepackage{amssymb}
\usepackage{txfonts}
\usepackage{float}
\usepackage{enumitem}
\usepackage{multirow}

\newcommand{\meq}[1]{(\ref{#1})}

\newcommand{\pp}{\partial}

\begin{document}

\title{Shadow of slowly rotating Kalb-Ramond black holes}

\author{Wentao Liu}
\affiliation{Department of Physics, Key Laboratory of Low Dimensional Quantum Structures and Quantum Control of Ministry of Education, and Synergetic Innovation Center for Quantum Effects and Applications, Hunan Normal
University, Changsha, Hunan 410081, P. R. China}
\author{Di Wu}
\email[]{wdcwnu@163.com}
\affiliation{School of Physics and Astronomy, China West Normal University, Nanchong, Sichuan 637002, P. R. China}
\author{Jieci Wang}
\email{jcwang@hunnu.edu.cn}\affiliation{Department of Physics, Key Laboratory of Low Dimensional Quantum Structures and Quantum Control of Ministry of Education, and Synergetic Innovation Center for Quantum Effects and Applications, Hunan Normal
University, Changsha, Hunan 410081, P. R. China}

\begin{abstract}

Real astronomical objects possess spin, yet deriving exact solutions for rotating black holes within gravitational theories is a formidable challenge.
To understand the shadow of rotating black holes in Lorentz-violating spacetimes induced by antisymmetric tensor fields, known as Kalb-Ramond (KR) fields, we have focused on the slow-rotation approximation framework.
Using this approach, we have obtained first-order rotation series solutions, which describe slowly rotating KR black holes.
For this solutions, we have plotted the black hole shadow contours under various parameters using the numerical backward ray-tracing method.
As the Lorentz-violating parameter increases, not only the apparent size of the black hole shadow decreases, but also the effects of rotation, such as the D-shaped structure and frame-dragging, are amplified.
Furthermore, the KR field also enhances gravitational lensing, causing the shadow to occupy a larger area within the photon ring.
This distinctive feature can differentiate KR gravity from general relativity.
Additionally, using the latest observational data from EHT on M87* and Sgr A*, we have provided constraints on the Lorentz-violating parameter of rotating KR black holes.
We found that, compared to static black holes, rotating black holes allow for the presence of stronger Lorentz violation effects.

\end{abstract}

\pacs{~}

\maketitle

\section{Introduction}

A black hole is akin to a coin placed beneath a sheet of paper; though invisible, its effect on light distribution is observable.
The shape and features of the coin are constant, and each time you rub the paper, the impression reveals the same characteristics.
Similarly, the spacetime structure of a black hole is fixed.
Regardless of the shape, color, or behavior of external electromagnetic radiation, the influence of the black hole remains constant as long as a few basic rules are followed.
The presence of the black hole can be unveiled through illumination from external electromagnetic sources.

Recently, the Event Horizon Telescope (EHT) Collaboration published the images of M87* \cite{EventHorizonTelescope:2019dse,EventHorizonTelescope:2019uob,EventHorizonTelescope:2019jan,EventHorizonTelescope:2019ths,EventHorizonTelescope:2019pgp,EventHorizonTelescope:2019ggy} and Sgr A* \cite{EventHorizonTelescope:2022wkp,EventHorizonTelescope:2022apq,EventHorizonTelescope:2022wok}, which are supermassive black holes.
These images reveal the effects of a constant spacetime structure illuminated by a time-varying emission region, urging us to better understand and interpret what we observe and what remains hidden.
A key feature of these images is the black hole shadow, whose shape and size provide a distinct signature of the celestial object.
Studying these shadows is crucial not only for identifying black holes but also for testing theories of gravity, including general relativity (GR) and modified gravity.
Moreover, it aids in addressing fundamental problems in physics, such as the extreme environments of black holes, the nature of dark matter, the behavior of accretion disks, the mechanisms driving an accelerating universe, and the potential existence of extra dimensions.
For further examples, see \cite{Grenzebach:2014fha,Abdujabbarov:2015xqa,Guo:2018kis,Zhu:2019ura,Long:2019nox,Chowdhuri:2020ipb,Lee:2021sws,Konoplya:2021slg,Zhang:2021hit,Nampalliwar:2021tyz,Junior:2021svb,Qin2022,Wang:2022kvg,Zeng:2021mok,Zhang:2022klr,Wang:2022mjo,Ghosh:2022mka,Vagnozzi:2022moj,Zubair:2023cor,Galishnikova:2023ltq,Broderick:2023jfl,Zhang:2024jrw,Jiang:2023img,Nguyen:2023clb,Huang:2024wpj,Wang:2023jop,Chen:2023wzv,Raza:2024zkp,Wei:2024cti,Liu:2024lbi,Kuang:2024ugn}.

Considering the quantum properties for modifications to gravitational theory is very important, as the study of Lorentz symmetry breaking (LSB) plays a crucial role in understanding gravity processes in fundamental physics \cite{Kostelecky1991,Kostelecky1998,Casana2018,Ovgun2019,Gullu2020,Pan:2020zbl,Liu:2022xse,Maluf2021,Xu:2022frb,Ding2022,Poulis:2021nqh,Mai:2023ggs,Xu:2023xqh,Zhang:2023wwk,Lin:2023foj,Chen:2023cjd,Chen2020,Wang:2021gtd,Liu:2024Lv,Mai:2024lgk,Liang:2022gdk,Tian:2021mka,Tian:2022gfa,Hosseinifar:2024wwe,Finke:2024ada,Liu:2024axg}.
By investigating the low-energy contributions from LSB, especially its impact on spacetime, and analyzing black hole shadows, we can explore the possibility of LSB in spacetimes compared to those predicted by GR.
A notable model of Lorentz violation is the Kalb-Ramond (KR) gravity theory, which includes an antisymmetric tensor field $B_{ab}$ , known as the KR field, that is nonminimally coupled to gravity. 
This tensor field originates from the bosonic spectrum of string theory \cite{Kalb:1974yc,Kao:1996ea}.
Within this framework, considerable research has been dedicated to discovering exact solutions \cite{Chakraborty:2014fva,Maluf:2021ywn,Duan:2023gng,Yang:2023wtu}.
Notably, Yang \textit{et al.} \cite{Yang:2023wtu} were the first to correctly derive the Schwarzschild-like solution within this theoretical framework.
They also obtained the Schwarzschild-(A)dS-like solution by relaxing the vacuum conditions.
Following this, we identified more general spherically symmetric neutral solutions in the same theory \cite{Liu:2024oas}.
Building on this discovery, the properties of these black holes have attracted considerable attention \cite{Guo:2023nkd,Filho:2023ycx,AraujoFilho:2024rcr,Jha:2024xtr,Junior:2024ety,Junior:2024vdk,Filho:2024kbq,Du:2024uhd,Filho:2024tgy}.

Given that real astronomical objects possess spin, it is crucial to consider the impact of angular momentum on these research findings.
However, obtaining exact solutions for rotating black holes within KR gravity presents a significant challenge and cannot be directly achieved by solving ordinary differential equations (ODEs) as in the static case or by using rotating solution generation techniques.
Yet, insights from the LIGO-Virgo Consortium indicate that most events detected during the $ O1 $ and $ O2 $ observations involve black holes with low effective spin values \cite{Deng:2024ayh}.
Roulet and Zaldarriaga conducted a thorough reanalysis of LIGO-Virgo strain data from 10 binary black hole mergers, examining models for angular spin distributions \cite{Roulet:2018jbe}.
They determined that typical black hole spins are generally limited to $ a/M\lesssim0.4 $, even when the spin orientations are random.
When the spins are aligned, the constraints become more stringent, with typical spins around $ a/M\sim 0.1 $.
A significant case is GW190814, which involved a merger between a $23 M_\odot $ black hole and a $ 2.6M_\odot $ compact object.
The large mass ratio in this event imposed a strict limit on the primary black hole's spin, constraining it to $ a/M\lesssim0.07 $ \cite{LIGOScientific:2020zkf}.
Moreover, relevant studies in black hole perturbation theory confirm the high accuracy of slow-rotation approximation framework \cite{Pani2011,Pani2013prl,Pani2013IJMPA,Liu:2022dcn}.
The real and imaginary parts of the QNMs in Kerr black holes between the slow-rotation approximation and the exact results are maintained within $1\%$ for this parameter range $ a/M\leq0.3 $ \cite{Pani2012}.
These discussions enable us to study the black hole shadow in scenarios involving slow rotations.

In this work, we aim to explore the impact of spontaneous Lorentz violation on the shadow contours of rotating black holes and constrain the Lorentz-violating parameters by numerical estimation of the angular radius of the supermassive black holes Sgr A* and M87* within this theoretical framework.
For this, we first extend the Schwarzschild-like solutions to include a rotation parameter $ a=J/M $ that describes the specific angular momentum of the black holes in slow-rotation approximation framework, thereby obtaining solwly rotating KR black holes.
Then, we evaluated the contour differences between the Kerr black hole shadow and the slowly rotating Kerr black hole shadow, obtaining the effective range of the black hole spin parameter under the slow rotation approximation framework.
Finally, we used the backward ray-tracing method to plot the shadow images of the slowly rotating KR black hole.

The rest of this paper is organized as follows.
In Sec. \ref{Sec.2}, we provide a concise overview of the spontaneous Lorentz symmetry breaking tensor field, i.e., the KR field, and present the slowly rotating KR black hole solutions for the Einstein-Kalb-Ramond gravitational field equations.
In Sec. \ref{Sec.3}, using the geodesic equation, we derive the orbital equations for photons in the context of slowly rotating KR spacetime.
In Sec. \ref{Sec.4}, we investigate the shadows of black holes, focusing on the apparent shapes and distortions of the shadows.
Additionally, we numerically calculate the observable measurements $R_s$ and $\delta_s$ to characterize the shadow.
Finally, Sec. \ref{Sec.5} provides a summary of our conclusions and discusses potential future research directions.

\section{Basic Formalism}\label{Sec.2}
\subsection{KR gravity theory}
Let us consider the Einstein-Hilbert action nonminimally coupled to a self-interacting KR field \cite{Altschul:2009ae}, as
\begin{equation}
\begin{aligned}\label{action}
\mathcal{S}=&\frac{1}{2\kappaup}\int d^4x\sqrt{-g}\bigg[R-2\Lambda-\frac{1}{6}H^{abc}H_{abc}-V(B^{ab}B_{ab})\\
&+\xi_2B^{ca}B^{b}{}_{a}R_{cb}+\xi_3B^{ab}B_{ab}R\bigg]+\int d^4x\sqrt{-g}\mathcal{L}_\text{M},
\end{aligned}
\end{equation}
where $ \kappaup=8\pi G_N $ is the gravitational coupling constant.
Here, $ \Lambda $ is the cosmological constant and $ \xi_2 $ and $ \xi_3 $ are the real coupling constants which constrain the nonminimal gravity interaction with the KR field.
$ H_{abc} $ is the totally antisymmetric field-strength tensor, defined by
\begin{align*}
H_{abc}\equiv\partial_aB_{bc}+\partial_b B_{ca}+\partial_c B_{ab}.
\end{align*}
The potential $ V $, chosen to ensure a nonzero VEV for the KR field and to be zero at its minimum, triggers spontaneous Lorentz symmetry breaking.
It is worth noting that the term $ \xi_3 B^{ab}B_{ab}R $ in the action \meq{action} transforms to $ \mp \xi_3 b^2 R $ in the vacuum.
This transformation allows the term to be absorbed into the Einstein-Hilbert action through a redefinition of variables.

By varying the action \eqref{action} with respect to the metric $ g_{ab} $, we obtain the following gravitational field equations:
\begin{align}
R_{ab}-\frac{1}{2}g_{ab}R+\Lambda g_{ab}=T^\text{KR}_{ab}+T^\text{M}_{ab},
\end{align}
where $ T^\text{M}_{ab} $ is the energy-momentum tensor of matter fields, and
\begin{equation}
\begin{aligned}
T^\text{KR}_{ab}=&\frac{1}{2}H_{acd}H_{b}{}^{cd} -\frac{1}{12}g_{ab}H^{cde}H_{cde}+2V'(X)B_{ca}B^{c}{}_{b} \\
&-g_{ab}V(X) +\xi_2\bigg[\frac{1}{2}g_{ab}B^{ce}B^{d}{}_{e}R_{cd} -B^{c}{}_{a}B^{d}{}_{b}R_{cd} \\
&-B^{cd}B_{bd}R_{ac} -B^{cd}B_{ad}R_{bc} +\frac{1}{2}\nabla_{c}\nabla_{a}(B^{cd}B_{bd}) \\
&+\frac{1}{2}\nabla_{c}\nabla_{b}(B^{cd}B_{ad}) -\frac{1}{2}\nabla^c\nabla_c(B_{a}{}^{d}B_{bd}) \\
&-\frac{1}{2}g_{ab}\nabla_{c}\nabla_{d}(B^{ce}B^{d}{}_{e})\bigg],
\end{aligned}
\end{equation}
which can be considered as the energy-momentum tensor of the KR field.
The prime indicates the derivative with respect to the variable of the corresponding functions.

Inspired by the gravitational sector of the Standard Model Extension, we consider a self-interacting potential for the KR field, which has a nonvanishing VEV, i.e., $ \langle B_{ab}\rangle =\beta_{ab} $ \cite{Altschul:2009ae}.
To achieve this, we assume the potential $V$ takes the following form
\begin{equation*}
V=V(B_{ab}B^{ab}\pm b^2),
\end{equation*}
where the sign $ \pm $ ensures that $ b^2 $ is a positive constant.
The VEV is then determined by the constant norm condition $ \beta^{ab}\beta_{ab}=\mp b^2 $, leading to the spontaneous breaking of Lorentz symmetry due to the self-interaction of the KR field \cite{Lessa:2019bgi}.
Additionally, the gauge invariance $ B_{ab}\rightarrow B_{ab}+\partial_{[a}\Gamma_{b]} $ of the KR field is spontaneously broken \cite{Yang:2023wtu}.
A prime example of a potential that meets these conditions is a smooth quadratic function, similar to that proposed by Casana et al. \cite{Casana2018}, given by:
\begin{align}
V=V(X)=\frac{1}{2}\lambda X^2,
\end{align}
where $\lambda$ is a constant and $X$ represents a generic potential argument.
Consequently, the VEV, $ \beta_{ab} $, arises as a solution of $ V=V'=0$.
However, this assumption is only applicable in the special case where the cosmological constant is not considered.
To explore the effect of Lorentz violation in asymptotically (A)dS spacetime with a nonzero cosmological constant, the vacuum conditions can be relaxed.
Similar to the potential assumed by Maluf \cite{Maluf2021}, another simple choice of potential is a linear function:
\begin{align}
V=V(\lambda,X)=\lambda X.
\end{align}
In this context, $ \lambda $ is interpreted as a Lagrange-multiplier field \cite{Bluhm:2007bd}.
The equation of motion for $\lambda$ ensures the vacuum condition $ X=0 $, resulting in $ V=0 $ for any on-shell field $ \lambda $.
Interestingly, the potential function in the above form behaves similarly to a cosmological constant. This particular assumption leads us to consider:
\begin{align}\label{VVV}
&V(\beta^{ab} \beta_{ab}+b^2)=\lambda(\beta^{ab}\beta_{ab}+b^2)=0,\\ \label{Vp}
&V'(\beta^{ab}\beta_{ab}+b^2)=\lambda,
\end{align}
where $ V'(X)=dV(X)/dX $.

Under the VEV configuration, i.e., $ B_{ab}B^{ab}=\beta_{ab}\beta^{ab} $, we can consider the exterior differential form \cite{Lessa:2019bgi}
\begin{align*}
\beta_2=-\tilde{E}(r)dt \wedge dr.
\end{align*}
The only nonvanishing terms are $ \beta_{rt}=-\beta_{tr}=\tilde{E}(r) $ in the VEV.
Equivalently, in terms of matrix forms:
\begin{equation*}
\beta_{ab}=\left(
\begin{array}{cccc}
0 & -\tilde{E}(r) & 0 & 0 \\
\tilde{E}(r) & 0 & 0 & 0 \\
0 & 0 & 0 & 0 \\
0 & 0 & 0 & 0
\end{array}\right),
\end{equation*}
it follows that the vacuum field exhibits a pseudoelectric configuration.
Consequently, this configuration automatically causes the KR field strength to vanish, i.e., $ H_{abc}=0 $ or $ H_3=d\beta_2=0 $ \cite{Yang:2023wtu}.

Then, we can define the efficient gravitational field equation, satisfying $\mathcal{G}_{ab}=0$, as follows:
\begin{align}\label{EQG}
\mathcal{G}_{ab}=R_{ab}-\Lambda g_{ab}-\xi_2\mathcal{B}_{ab}-(2\beta_{ac} \beta_{b}{}^{c}+b^2g_{ab})\lambda,
\end{align}
with
\begin{equation}\label{EQB}
\begin{aligned}
\mathcal{B}_{ab} = &g_{ab}\beta^{ce}\beta^{d}{}_{e}R_{cd} -\beta^{c}{}_{a}\beta^{d}{}_{b}R_{cd} -\beta^{cd}\beta_{ad}R_{bc} \\
&-\beta^{cd}\beta_{bd}R_{ac} +\frac{1}{2}\nabla_c\nabla_{a}(\beta^{cd}\beta_{bd}) \\
&+\frac{1}{2}\nabla_c\nabla_{b}(\beta^{cd}\beta_{ad}) -\frac{1}{2}\nabla^c\nabla_c(\beta_a{}^{d}\beta_{bd}).
\end{aligned}
\end{equation}
The covariant derivative can be expanded using the Christoffel symbols, as detailed in appendix A of Ref. \cite{Liu:2024oas}.

\subsection{Slowly rotating KR black holes}
In this subsection, we assume the metric corresponds to a slowly rotating black hole and adopt the following line element \cite{Pani2012}:
\begin{equation}
ds^2= -A(r)dt^2+B(r)dr^2+r^2d\Omega^2-2a\varpi(r,\theta)dt d\varphi,
\end{equation}
where $ d\Omega^2=d\theta^2+\sin^2\theta d\varphi^2 $ and $ a $ is a parameter associated with its angular momentum.
Here and hereafter we linearize all quantities in the spin parameter $ \tilde{a}\equiv a/M\equiv J/M^2 $, and neglecting terms of high-order $ \mathcal{O}(\tilde{a}^2) $.
The form corresponding to the above metric for the pseudoelectric field $ \tilde{E}(r) $ is given by:
\begin{align}
\tilde{E}(r)=|b|\sqrt{\frac{A(r)B(r)}{2}},
\end{align}
so that the constant norm condition $ \beta^{ab}\beta_{ab}=-b^2 $ is satisfied.
Then, the nonzero components of the efficient gravitational field equations, associated with the metric, are
\begin{align} \label{E00}
&\mathcal{G}_{tt}=\frac{(1-\ell)A}{4rB}\left(4A'/A-r\varUpsilon \right)+\Lambda~A+\mathcal{O}(\tilde{a}^2),\\  \label{E11}
&\mathcal{G}_{rr}=\frac{(1-\ell)}{4r}\left(4B'/B+r\varUpsilon \right)-\Lambda~B+\mathcal{O}(\tilde{a}^2), \\
&\begin{aligned} \label{E22}
\mathcal{G}_{\theta\theta}=&\frac{1}{2B}\left[\ell r^2\varUpsilon-r(1+\ell)(A'/A-B'/B)+2\ell-2  \right]\\
&+1-r^2b^2\lambda-r^2\Lambda+\mathcal{O}(\tilde{a}^2),
\end{aligned}\\
&\begin{aligned}\label{G03}
\mathcal{G}_{t\varphi}=&\frac{(1-4\ell)\tilde{a}M}{2B}\left[2\pp^2_r\varpi-\pp_r(AB)/(AB)\pp_r\varpi  \right]\\
&+\frac{(1-\ell)\tilde{a}M}{2r^2}\left[\pp^2_\theta\varpi-\cot\theta \pp_\theta \varpi \right]\\
&+\frac{l\tilde{a}M}{2r^2B}\left[2-2B-rA'/A-3rB'/B-r^2\varUpsilon \right]\varpi\\
&+\frac{\tilde{a}M}{r}\left[ A'/(AB)+rb^2\lambda+r\Lambda  \right]\varpi+\mathcal{O}(\tilde{a}^2),
\end{aligned}
\end{align}
and $ \mathcal{G}_{\varphi t}=\mathcal{G}_{t\varphi} $, $ \mathcal{G}_{\varphi\varphi}=\mathcal{G}_{\theta\theta}\sin^2\theta $,
where $ \varUpsilon \equiv(A'/A)^2-2A''/A+A'B'/(AB) $ and $ \ell\equiv \xi_2 b^2/2 $.
If $ \tilde{a}=0 $, Eq. \meq{G03} vanishes.

Combining Eqs. \eqref{E00} and \eqref{E11}, we obtain $ \mathcal{G}_{tt}+\frac{A}{B}\mathcal{G}_{rr}=0 $. This leads to the first constraint condition, as follows:
\begin{align}\label{BAB}
\partial_r(AB)+\mathcal{O}(\tilde{a}^2)=0 \Rightarrow B(r)=\frac{\mathcal{C}_1}{A(r)},
\end{align}
where $ \mathcal{C}_1 $ is an arbitrary constant.
Therefore, the relation $ G_{\theta\theta}=0 $ gives the following equation involving only the metric function $A(r)$:
\begin{align}\label{ApdL}
\left(\ell \pp_r^2+\frac{1+\ell}{r}\pp_r+\frac{1-\ell}{r^2}\right)A(r)-\frac{\mathcal{C}_1}{r^2}(1-r^2b^2\lambda-r^2\Lambda)=0.
\end{align}
The above equation can be solved to obtain:
\begin{align}\label{caseAB}
A(r)=\frac{\mathcal{C}_1}{1-\ell}+\frac{\mathcal{C}_2}{r}+\frac{\mathcal{C}_3}{r^{n}}-\frac{\mathcal{C}_1\left(b^2\lambda+\Lambda\right)}{3(1+\ell)}r^2,
\end{align}
where $ \mathcal{C}_2 $ and $ \mathcal{C}_3 $ are integration constants, and $ n $ is determined by the Lorentz-violating parameter as $ \frac{1-\ell}{\ell}$.
The constraint relation of the solution \meq{caseAB} for the equations $ G_{tt}=G_{rr}=0 $ can be simplified to
\begin{align}\label{C123}
\frac{\mathcal{C}_3n(n-1)}{r^{n+2}}+\frac{4\mathcal{C}_1}{(1+\ell)}\left(\frac{\ell \Lambda}{1-\ell}-\frac{b^2\lambda}{2}\right)=0.
\end{align}
To maintain consistency with the power exponent terms of the solutions in Ref. \cite{Lessa:2019bgi}, it is necessary to satisfy $ n=-2 $, $ n=0 $, or $ n=1 $.
In any case, this term will be absorbed by other terms; therefore, in our present work, it is evident that $ \mathcal{C}_3=0 $.
Then, the constraint relation \eqref{C123} can be simplified to
\begin{align}\label{eLl}
2\ell \Lambda-(1-\ell)b^2\lambda=0.
\end{align}
The above equation is a necessary condition for generating a solution with a cosmological constant in the KR framework due to the characteristics of the effective gravitational field equation \eqref{EQG} and the potential \eqref{Vp} \cite{Maluf2021,Yang:2023wtu}.
Using the constraint \eqref{eLl}, one can introduce an effective cosmological constant $ \Lambda_e $ to replace $ \Lambda $ and $ \lambda $, as follows:
\begin{align}
\Lambda = (1-\ell)\Lambda_e,~~~~~~ \lambda = \frac{2\ell}{b^2} \Lambda_e.
\end{align}

To examine the effect of rotation, we derive a second-order partial differential equation (PDE) for the gyromagnetic function $ \varpi(r,\theta) $ from the condition $ \mathcal{G}_{t\varphi}=0 $ as follows
\begin{align}\label{PDEs}
\tilde{a}\left[\hat{\mathcal{D}}_r+(1-\ell)\hat{\mathcal{D}}_\theta\right]\varpi(r,\theta)+\mathcal{O}(\tilde{a}^2)=0.
\end{align}
Here, the operators $ \mathcal{D}_r $ and $ \mathcal{D}_\theta $ are defined by
\begin{equation}
\begin{aligned}
\hat{\mathcal{D}}_r=F(1-2\ell)/\mathcal{C}_1(r^2\pp^2_r-2),~~
\hat{\mathcal{D}}_\theta=\pp^2_\theta-\cot\theta \pp_\theta+2,
\end{aligned}
\end{equation}
with
\begin{align}\label{FFF}
F(r)=A(r)|_{\mathcal{C}_3\rightarrow0}=\frac{\mathcal{C}_1}{1-\ell}+\frac{\mathcal{C}_2}{r}-\frac{\mathcal{C}_1\Lambda_e}{3}r^2.
\end{align}
In order to solve the PDE \eqref{PDEs}, it is necessary to separate variables for the gyromagnetic function $ \varpi(r,\theta) $. Since
\begin{align}
\mathcal{D}_\theta \sin^2\theta=0,
\end{align}
one can set $ \varpi(r,\theta)=\sin^2\theta \varpi(r) $ to eliminate the angular part in the PDE, which becomes an ODE, as
\begin{align}\label{ODEs}
\tilde{a}\hat{\mathcal{D}}_r\varpi(r)+\mathcal{O}(\tilde{a}^2)=0.
\end{align}
The solution of the ODE \meq{ODEs} is
\begin{align}
\varpi(r)=\frac{\mathcal{C}_4}{r}+\mathcal{C}_5 r^2.
\end{align}
Now, All constants must be determined to obtain a solution that describes black hole spacetimes.
The asymptotic behavior of asymptotically (A)dS slowly rotating vacuum solutions is identical to that of slowly rotating Kerr-(A)dS solutions; thus, $ \mathcal{C}_5 $ can be set to $ \mathcal{C}_1\Lambda_e/3 $.
Then, the gyromagnetic function $ \varpi(r,\theta) $ will reduce to the slowly rotating Kerr solution when $ \ell=0 $, therefore, one can set $ \mathcal{C}_4=2 M$.
The constant $ \mathcal{C}_2 $ can be interpreted as the black hole mass parameter with $ \mathcal{C}_2=-2M $.
To ensure that our solutions reduce to the Schwarzschild solution when $ \ell=\tilde{a}=0 $, the constant $ \mathcal{C}_1 $ has two choices, referred to as Case A and Case B \cite{Liu:2024oas}.

\textbf{Case A:}
Firstly, we consider a general spherically symmetric structure.
By setting $ \mathcal{C}_1=1-\ell $, we can rewrite Eq. \meq{FFF} as follows:
\begin{align}
F_\text{A}(r)=1-\frac{2M}{r}-\frac{(1-\ell)\Lambda_e}{3}r^2,
\end{align}
which is the same as that of the Schwarzschild-(A)dS case.
However, the metric function $ B(r)=(1-\ell)F(r)^{-1} $ ensures that the spacetime structure differs from that of a slowly rotating Kerr-(A)dS black hole.
Thus, the slowly rotating Kerr-KR-(A)dS metric of Case A is:
\begin{equation}
\begin{aligned}\label{caseA}
ds_A^2=&-F_\text{A}(r)dt^2+\frac{(1-\ell)}{F_\text{A}(r)}dr^2+r^2d\theta^2+r^2\sin^2\theta d\varphi^2\\
&-2\tilde{a}M\left(\frac{2M}{r}+\frac{(1-\ell)\Lambda_e}{3}r^2\right)\sin^2\theta dtd\varphi+\mathcal{O}(\tilde{a}^2).
\end{aligned}
\end{equation}
In the first-order slow-rotation limit, the ring singularity is degenerated into a point singularity and the Cauchy horizon is vanished, as the terms responsible for their distinction are considered higher-order perturbations and thus negligible.
The horizons are determined by the equation $ F_\text{A}(r)=0 $, which can be expressed as follows:
\begin{equation}
\begin{aligned}
F_\text{A}^\text{dS}&=(1-\ell)\frac{\Lambda_e}{3}\left(1-\frac{r_h}{r}\right)\left(r_c-r\right)\left(r+r_h+r_c \right),\\
F_\text{A}^{\text{AdS}}&=\left(1-\frac{r_h}{r}\right)\left[1-(1-\ell)\frac{\Lambda_e}{3}(r^2+r_hr+r_h^2)\right].
\end{aligned}
\end{equation}
In asymptotically AdS spacetime, there is only an event horizon located at $r = r_h$, while in asymptotically dS spacetime, there is an additional cosmological horizon located at $r = r_c$.

\textbf{Case B:}
In the following, we consider a special spherically symmetric structure.
By setting $ \mathcal{C}_1=1 $, one can rewrite Eq. \meq{FFF} as follows:
\begin{align}
F_\text{B}(r)=\frac{1}{1-\ell}-\frac{2M}{r}-\frac{\Lambda_e}{3}r^2,
\end{align}
and using the relation in Eq. \meq{BAB}, it is easy to obtain the slowly rotating Kerr-KR-(A)dS metric of Case B as
\begin{equation}\label{caseB}
\begin{aligned}
ds^2=&-F_\text{B}(r)dt^2+\frac{1}{F_\text{B}(r)}dr^2+r^2d\theta^2+r^2\sin^2\theta d\varphi^2\\
&-2\tilde{a}M\left(\frac{2M}{r}+\frac{\Lambda_e}{3}r^2\right)\sin^2\theta dtd\varphi+\mathcal{O}(\tilde{a}^2).
\end{aligned}
\end{equation}
The function $ F_\text{B}(r) $ can be expressed in terms of the black hole horizons as follows:
\begin{equation}
\begin{aligned}
F_\text{B}^\text{dS}&=\frac{\Lambda_e}{3}\left(1-\frac{r_h}{r}\right)\left(r_c-r\right)\left(r+r_h+r_c \right),\\
F_\text{B}^{\text{AdS}}&=\left(1-\frac{r_h}{r}\right)\left[\frac{1}{1-\ell}-\frac{\Lambda_e}{3}(r^2+r_hr+r_h^2)\right].
\end{aligned}
\end{equation}

In both Case A and Case B, the dimensionless parameters $ \ell $ and $ \tilde{a} $ represent the Lorentz violation parameter in spacetime and the spin parameter of the black holes, respectively.
When the Lorentz violation parameter $ \ell $ vanishes, the solutions reduce to the slowly rotating Kerr-(A)dS black hole cases \cite{Tattersall2018}; when $\tilde{a}=0$, they reduce to the static neutral KR black hole cases \cite{Liu:2024oas,Yang:2023wtu}.
In asymptotically flat slowly rotating vacuum cases, i.e., $ \Lambda_e=0 $, the event horizons in both Case A and Case B are $ r_h=2M $ and $ r_h=2(1-\ell)M $, respectively.

\section{Photon orbits}\label{Sec.3}

In this section, we provide a brief overview of photon trajectories in the gravitational field of slowly rotating KR spacetimes.
When an observer views a background source through the gravitational field of a black hole situated in between, the black hole's shadow contour emerges, formed by photons deflected by the black hole and not reaching the observer.
This highlights the photon's geodesics within the black hole spacetime.
The geodesics corresponding to photon geometry are defined by the Hamilton-Jacobi equation:
\begin{align}\label{EqHJ}
\frac{\pp \mathcal{S}}{\pp \tau}=-\frac{1}{2}g^{ab}\frac{\pp \mathcal{S}}{\pp x^a}\frac{\pp \mathcal{S}}{\pp x^b},
\end{align}
where $ \tau $ is the affine parameter of the null geodesic and $ \mathcal{S} $ denotes the Jacobi action of the photon, which can be separated into the following form,
\begin{align}\label{actionS}
\mathcal{S}=\frac{1}{2}m^2\lambda-\mathcal{E}t+L\varphi+\mathcal{S}_r(r)+\mathcal{S}_\theta(\theta).
\end{align}
Here, $ m $ denotes the mass of the particle moving in the black hole spacetime, and for a photon, we have $ m=0 $.
$ \mathcal{E} $ and $ L_z $ represent the energy and angular momentum of the photon in the direction of the rotation axis, respectively.
The functions $ \mathcal{S}_r(r) $ and $ \mathcal{S}_\theta(\theta) $ depend only on $ r $ and $ \theta $, respectively.

Utilizing the properties of Killing vector fields, two conserved quantities, $ \mathcal{E} $ and $ L_z $, can be defined as \cite{Waldbook,Liu2023}:
\begin{align}\label{gabE}
\mathcal{E}=&-g_{ab}\xi^a\dot{x}^b=-g_{tt}\dot{t}-g_{t\varphi}\dot{\varphi},\\ \label{gabL}
L_z=&g_{ab}\psi^a\dot{x}^b=g_{t\varphi}\dot{t}+g_{\varphi\varphi}\dot{\varphi}.
\end{align}

By combining Eqs. \meq{EqHJ}-\meq{gabL}, two equations of motion for propagating photons can be calculated as:
\begin{align}\label{taut}
\dot{t}=&\frac{\mathcal{E}}{A(r)}-\frac{\tilde{a}ML_z\varpi(r)}{r^2A(r)}+\mathcal{O}(\tilde{a}^2),\\ \label{tauvarphi}
\dot{\varphi}=&-\frac{L_z}{r^2 \sin^2\theta}-\frac{\tilde{a}M\mathcal{E}\varpi(r)}{r^2A(r)}+\mathcal{O}(\tilde{a}^2).
\end{align}
Substituting Eqs. \meq{taut}-\meq{tauvarphi} into the null geodesics equiation $ g_{ab}\dot{x}^a\dot{x}^b=0 $, one can obtain
\begin{equation}
\begin{aligned}\label{nge}
B(r)\dot{r}^2+r^2\dot{\theta}^2&+\frac{L_z^2\csc^2\theta}{r^2}-\frac{\mathcal{E}^2}{A(r)}\\
&+\frac{2\tilde{a}ML_z\mathcal{E}\varpi(r)}{r^2A(r)}+\mathcal{O}(\tilde{a}^2)=0.
\end{aligned}
\end{equation}
By employing the Carter constant $ \mathcal{K} $ \cite{Carter:1968rr} to separate variables in the above Eq. (\ref{nge}), two additional equations of motion for photons can be easily computed as
\begin{align}\label{taur}
r^2\dot{r}=&\sqrt{\mathcal{R}(r)},\\ \label{tauvartheta}
r^2\dot{\theta}=&\sqrt{\Theta(\theta)},
\end{align}
and the expressions of the radial and angular equations are as follows:
\begin{align}
\mathcal{R}=&\frac{r^4\mathcal{E}^2-2r^2\tilde{a}ML_z\mathcal{E}\varpi(r)}{A(r)B(r)}-\frac{(\mathcal{K}+L_z^2)r^2}{B(r)}+\mathcal{O}(\tilde{a}^2),\\
\Theta=&\mathcal{K}-L_z^2\cot^2\theta+\mathcal{O}(\tilde{a}^2).
\end{align}
These Eqs. \meq{taut}, \meq{tauvarphi}, \meq{taur} and \meq{tauvartheta} describe the propagation of light in spacetime around a slowly rotating KR black hole.

To study the trajectories of photons, it is convenient to write the radial geodesics in terms of the effective potential $ V_\text{eff}(r) $ as
\begin{align}\label{Veff}
\left(\frac{\sqrt{\mathcal{C}_1}}{\mathcal{E}}\frac{dr}{d\tau}\right)^2+V_\text{eff}(r)=0,
\end{align}
with
\begin{equation}
\begin{aligned}
V_\text{eff}(r)&=-r^{-4}\mathcal{R}(r)\mathcal{C}_1/\mathcal{E}^2
\\&=-1+\frac{A(r)(\eta+\xi^2)}{r^2}+\frac{4M^2\tilde{a}\xi}{r^3}+\mathcal{O}(\tilde{a}^2),
\end{aligned}
\end{equation}
where the symbol $\mathcal{C}_1=A(r)B(r)$ is a constant, which determines whether the black hole is Case A or Case B.
The impact parameters $\xi=L/\mathcal{E}$ and $\eta=\mathcal{K}/\mathcal{E}^2$ can be defined \cite{Chandrasekhar}.
We can consider the effective potential as a correction to the static case due to the black hole's spin \cite{AraujoFilho:2024rcr,Junior:2024ety,Jha:2024xtr}.

The black hole's shadow silhouette is derived from the specific orbit in the radial equation, defined by $r=r_p$, which satisfies the conditions \cite{Meng:2023uws}
\begin{align}\label{YGDTJ}
V_\text{eff}(r)\big|_{r=r_p}=0,&&\frac{d}{dr}V_\text{eff}(r)\big|_{r=r_p}=0.
\end{align}

We observe that the equations governing the shadow of the black hole depend only on the metric function $A(r)$ and are independent of $B(r)$.
This implies that in Case A, under the slow rotation approximation, Lorentz violation does not affect the characterization of the black hole's shadow, though potential effects may arise in higher-order rotational cases.

Then, by solving the system of equations \eqref{YGDTJ} for Case B, we obtain the resulting impact parameters that provide the necessary information:
\begin{align}
\xi=&\frac{r^3_p-3(1-\ell)Mr_p^2}{2(r^2_p\Lambda-1)M^2\tilde{a}},\\
\eta=&\frac{3(1-\ell)r^2_p}{1-r^2_p\Lambda}-\xi^2.
\end{align}
When $ \ell=0 $, it corresponds to the impact parameters in Case A or a slowly rotating Kerr-(A)dS black hole.
The two impact parameters determine the boundary of the black hole shadow; for simplicity, we set \(M=1\) for the black hole's mass in subsequent calculations.

\section{SHADOWS OF BLACK HOLES} \label{Sec.4}

Currently, the asymptotically flat rotating Kerr-like black hole is the most widely accepted model in astronomical observations.
Consequently, this section primarily focuses on black hole shadows in asymptotically flat spacetime.
In this section, we will to adopt the numerical backward ray-tracing method \cite{Wang:2017qhh,Wang:2021ara,Hu:2020usx,Zhong:2021mty,Chen:2023wna} to investigate the shadow of slowly rotating black holes.

\subsection{Apparent shape}\label{Sec41}
Usually, photons emitted by a light source are deflected when they pass near a black hole due to gravitational lensing effects \cite{Liu:2024bre}.
Some of the photons, after being deflected by the black hole, can reach a distant observer, while others fall directly into the black hole.
The photons that cannot escape form the black hole's shadow in the observer's sky.
We assume that an observer is located at $ (r_0,\theta_0) $ in the coordinates $ \{t,r,\theta,\varphi\} $.
In this case, we choose to use the following normalized and orthogonal tetrad \cite{Zhang:2020xub}:

\begin{equation}\label{zjbj}
\begin{aligned}
e_{(t)}=&\left.\sqrt{\frac{g_{\varphi\varphi}}{g_{t\varphi}^2-g_{tt}g_{\varphi\varphi}}}\left(\pp_t-\frac{g_{t\varphi}}{g_{\varphi\varphi}}\pp_\varphi\right)\right|_{(r_0,\theta_0)},\\
e_{(r)}=&\left.\frac{1}{\sqrt{g_{rr}}}\pp_r\right|_{(r_0,\theta_0)},\\
e_{(\theta)}=&\left.\frac{1}{\sqrt{g_{\theta\theta}}}\pp_\theta\right|_{(r_0,\theta_0)},\\
e_{(\varphi)}=&\left.\frac{1}{\sqrt{g_{\varphi\varphi}}}\pp_\varphi\right|_{(r_0,\theta_0)}.
\end{aligned}
\end{equation}
In this context, $ g_{ab} $  represents the metric component of the background spacetime.
It is important to note that this tetrad is not unique; one can select the appropriate tetrad based on specific requirements.
In a real physical scenario, light travels from the source to the observer.
However, for the sake of calculations, we can assume that the light originates from the observer because optical paths are reversible.

Using the orthogonal tetrad, the four-momentum can be expressed as follows:
\begin{align}\label{Meq}
p^{(t)}&=-p_\mu e^{\mu}_{(t)},\\
p^{(i)}&=p_\mu e^{\mu}_{(i)},
\end{align}
where $ i=r,\theta,\varphi $.
This represents the four-momentum as measured by a locally static observer.
Since the photon is massless, the three-vector linear momentum $ \vec{P} $ is related to  $ p^{(i)} $ and satisfies $ |\vec{P}|=p^{(t)} $ in the observer's frame.
The observation angles $ (\alpha,\beta) $ can be introduced as follows:
\begin{align}
p^{(r)}&=|\vec{P}|\cos\alpha \cos\beta,\\
p^{(\theta)}&=|\vec{P}|\sin\alpha,\\
p^{(\varphi)}&=|\vec{P}|\cos\alpha\sin\beta.
\end{align}

Actually, the angular coordinates $ (\alpha,\beta) $ of a point in the observer's local sky specify the direction of the corresponding light ray and establish its initial conditions.
The coordinates $ (X,Y) $ of a point in the local sky of the observer are related to its angular coordinates $ (\alpha,\beta) $ by:
\begin{align}
X&=-r_0\tan\beta=-r_0\frac{p^{(\varphi)}}{p^{(r)}},\\
Y&=r_0\frac{\tan\alpha}{\cos\beta}=r_0\frac{p^{(\theta)}}{p^{(r)}}.
\end{align}
The image of a black hole shadow in observer's sky is composed of the pixels corresponding the light rays falling down into the black hole horizon.
The unstable spherical orbits of photons provide us the boundary of the shadow.
In the slowly rotating Kerr black hole spacetime, the position of the image of the photon in the sky as seen by the observer is given by:
\begin{align}
X &= -r_0 \frac{p^{(\varphi)}}{p^{(r)}} = -\frac{r_0^3}{B(r_0)} \frac{L_z \sqrt{g_{rr}}}{\sqrt{g_{\varphi\varphi} \mathcal{R}(r_0)}}, \\
Y &= r_0 \frac{p^{(\theta)}}{p^{(r)}} = \frac{r_0^3}{B(r_0)} \frac{\sqrt{g_{rr} \Theta(\theta_0)}}{\sqrt{g_{\theta\theta} \mathcal{R}(r_0)}},
\end{align}
assuming the observer is located at a distance $r = r_0$ and $\theta = \theta_0$.
For a real observer located far from the asymptotically flat black hole in Case B, take the limit $r_0 \to \infty$, which yields:
\begin{align}\label{XXX}
X &= -\xi \csc \theta_0/\sqrt{1-\ell}, \\ \label{YYY}
Y &= \sqrt{\eta - \xi^2 \cot^2 \theta_0}/\sqrt{1 - \ell}.
\end{align}
When $ \ell=0 $, it corresponds to the slowly rotating Kerr case, which can also represent Case A.
It is easy to observe that the two celestial coordinates satisfy:
\begin{align}
\xi^2+\eta=(1-\ell)(X^2+Y^2),
\end{align}
which indicates that the Lorentz violation influences not only the modification of the impact parameters but also the overall scaling of the photon sphere radius.
Notably, for Case A, the coordinates $X$ and $Y$ remain the same as those for a slowly rotating Kerr black hole, implying that Lorentz violation does not affect the black hole image.
This conclusion is consistent with Eq. \meq{Veff}.

\begin{figure}[H]
\centering
\includegraphics[width=0.48\linewidth]{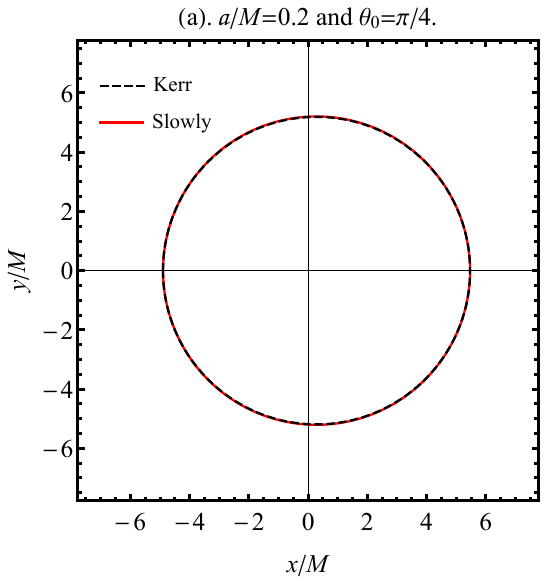}
\includegraphics[width=0.48\linewidth]{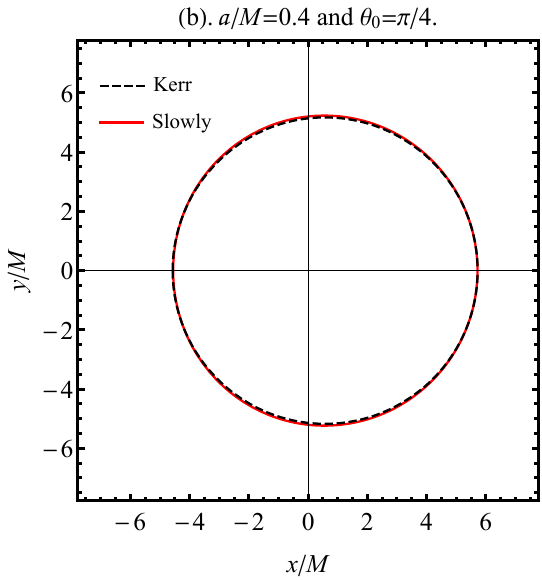}
\includegraphics[width=0.48\linewidth]{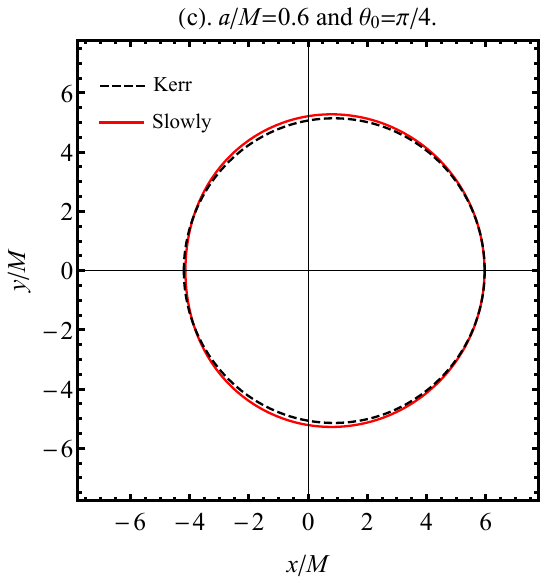}
\includegraphics[width=0.48\linewidth]{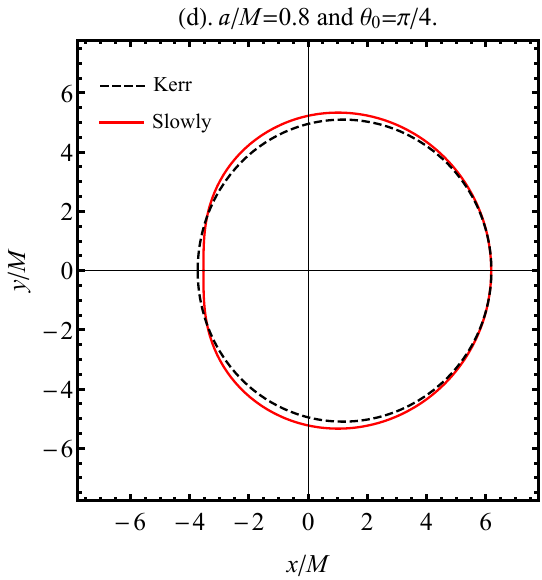}
\caption{The shadow contours of Kerr black holes are compared with those of slowly rotating Kerr black holes, which can also represent Case A.
Both cases are observed at an angle of $\pi/4$.}
\label{fig1}
\end{figure}
\begin{figure}[ht]
\centering
\includegraphics[width=0.48\linewidth]{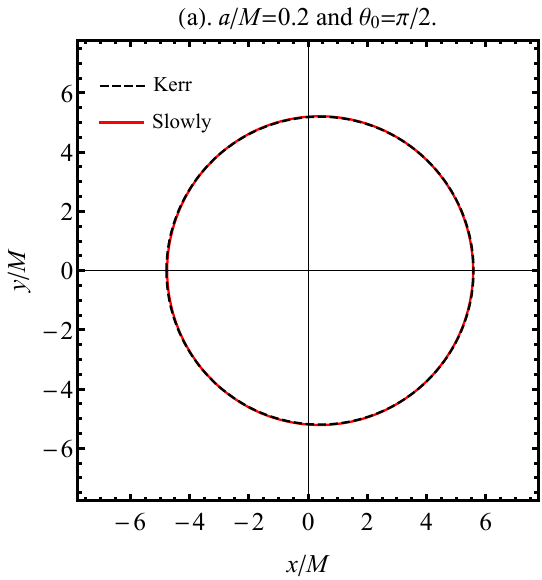}
\includegraphics[width=0.48\linewidth]{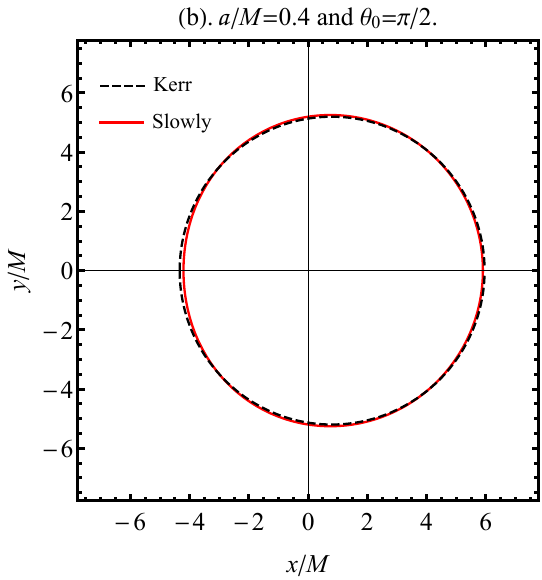}
\includegraphics[width=0.48\linewidth]{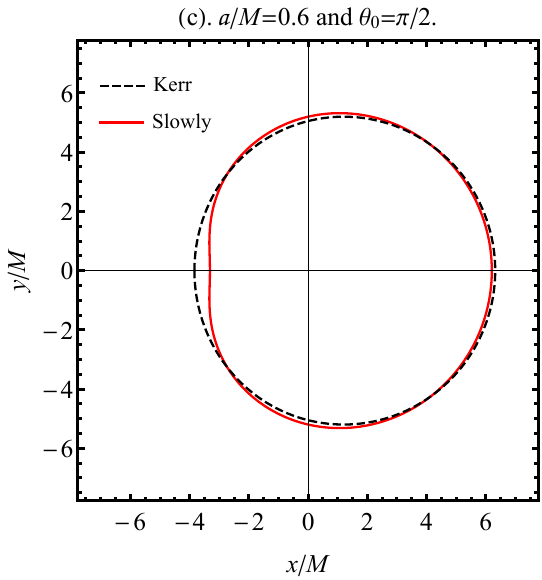}
\includegraphics[width=0.48\linewidth]{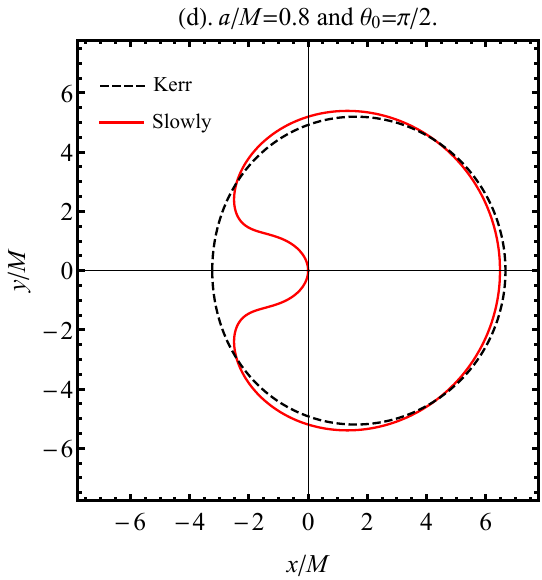}
\caption{The shadow contours of Kerr black holes are compared with those of slowly rotating Kerr black holes, which can also represent Case A.
Both cases are observed at an angle of $\pi/2$.}
\label{fig2}
\end{figure}

Before discussing the impact of Lorentz violation on the black hole shadow, we need to first discuss the parameter range applicable to the slow rotation approximation.
In existing studies of black hole perturbation theory \cite{Pani2013IJMPA}, the discrepancy between the slow-rotation approximation and the exact results is kept within $1\%$ for the parameter range $ (\tilde{a}\leq0.3) $.
In this section, we aim to provide a reasonable range for the spin parameter by comparing the shadow contours of Kerr black holes with those of slowly rotating Kerr black holes.

\begin{figure}[ht]
\centering
\includegraphics[width=0.85\linewidth]{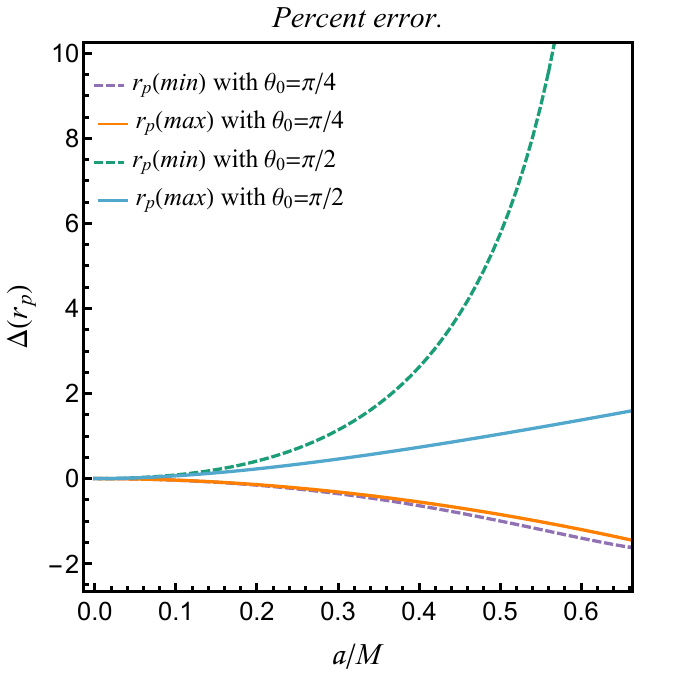}
\caption{The percent error between the innermost stable circular orbit and the outermost stable circular orbit of slowly rotating Kerr black holes (Case A) and Kerr black holes at different observation angles increases as the black hole spin parameter $ a/M $ increases.}
\label{fig3}
\end{figure}

In Figs. \ref{fig1}(a)-\ref{fig1}(d), we show the shadow contours for Kerr black holes and slowly rotating black holes at an observation angle of $\theta_0 = \pi/4$, which have nearly the same characteristic shape for $ a/M\leq0.6 $.
However, as the spin parameter approaches extremality, the shadow contours under the slow rotation approximation appear more bulging.
Figs. \ref{fig2}(a)-\ref{fig2}(d) show the comparison of shadow contours for different spin parameters at an observation angle of $\theta_0 = \pi/2$. Within the range of the spin parameter $a/M \leq 0.4$, the images fit very well.
The slow rotation approximation accurately reflects various characteristics of Kerr black holes.
For higher spin parameters, the D-shaped structure appears earlier, and in near-extremal cases, a heart-shaped shadow emerges.
We consider this to be due to the parameters exceeding the limits of the slow rotation approximation, resulting in mistakes.
This can be attributed to the rapid increase in error as the photon sphere radius grows with the spin parameter, as shown in Fig. \ref{fig3}.
Here, the percentage error of the ohoton sphere radius, for example, is defined as
\begin{align}
\Delta(r_p)=\frac{r_p^\text{Kerr}-r_p^\text{Slowly}}{r_p^\text{Slowly}}\times 100\%.
\end{align}

\begin{figure*}[ht]
\centering
\includegraphics[width=0.22\linewidth]{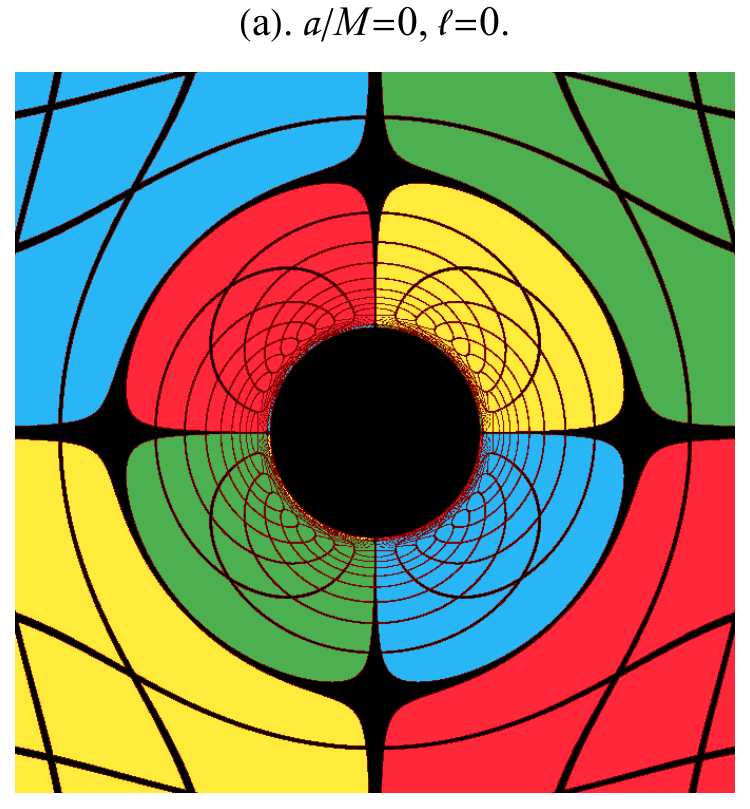}
\includegraphics[width=0.22\linewidth]{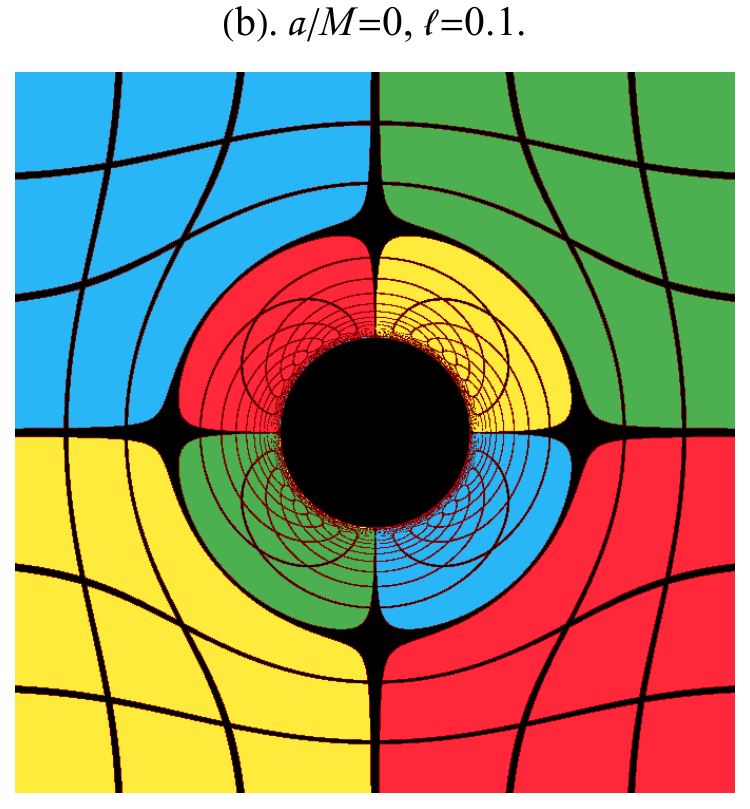}
\includegraphics[width=0.22\linewidth]{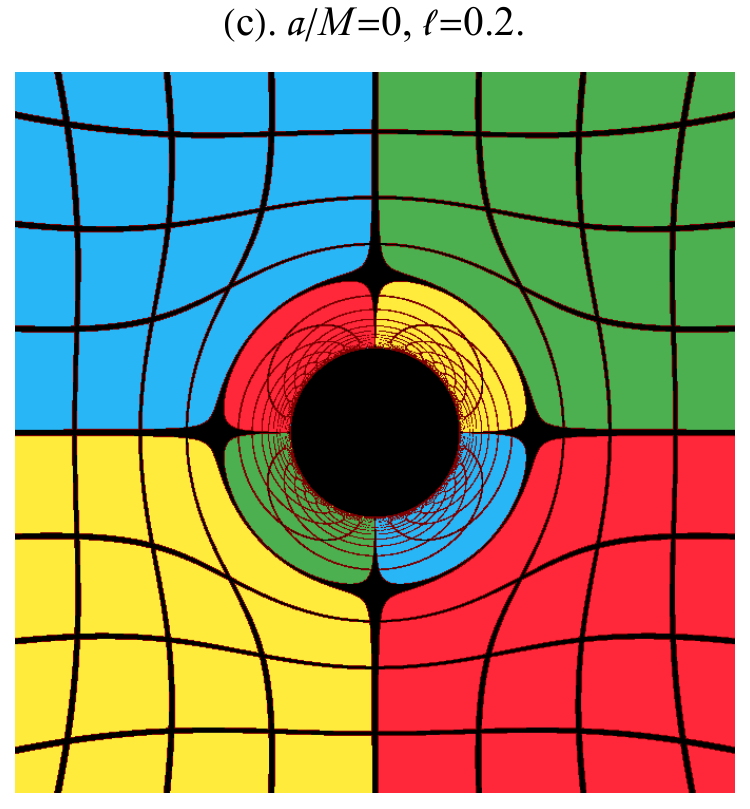}
\includegraphics[width=0.22\linewidth]{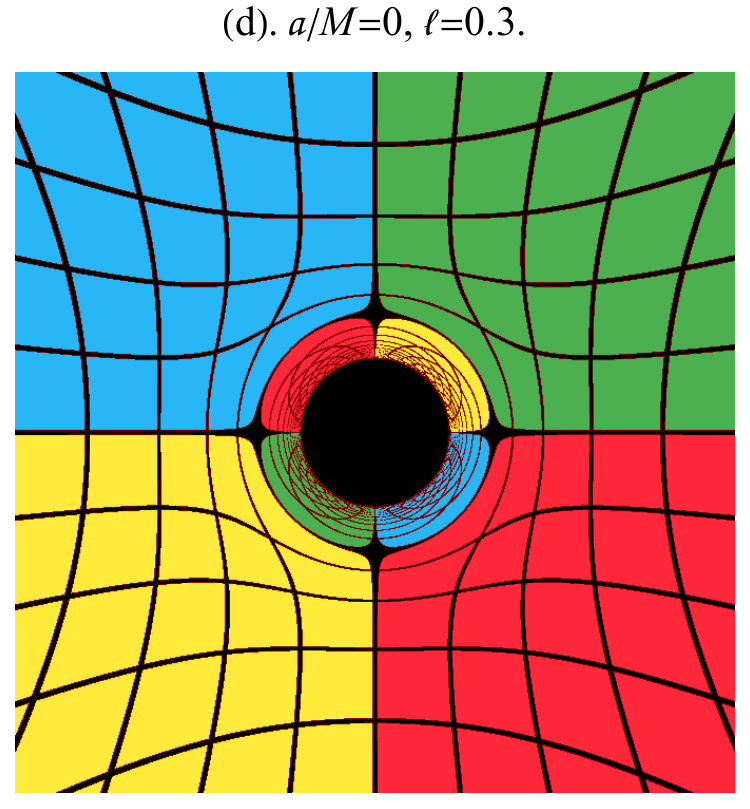}
\includegraphics[width=0.22\linewidth]{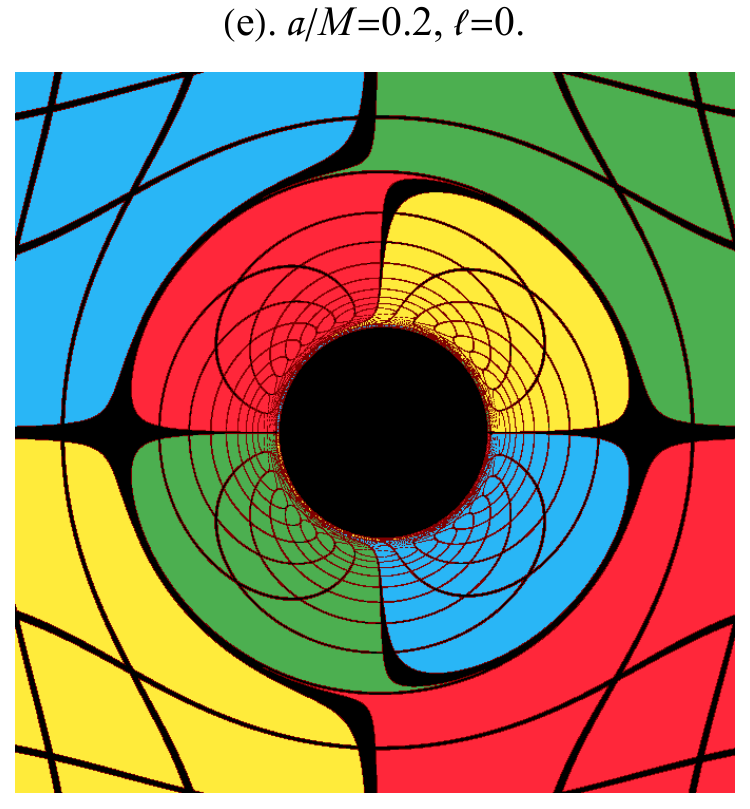}
\includegraphics[width=0.22\linewidth]{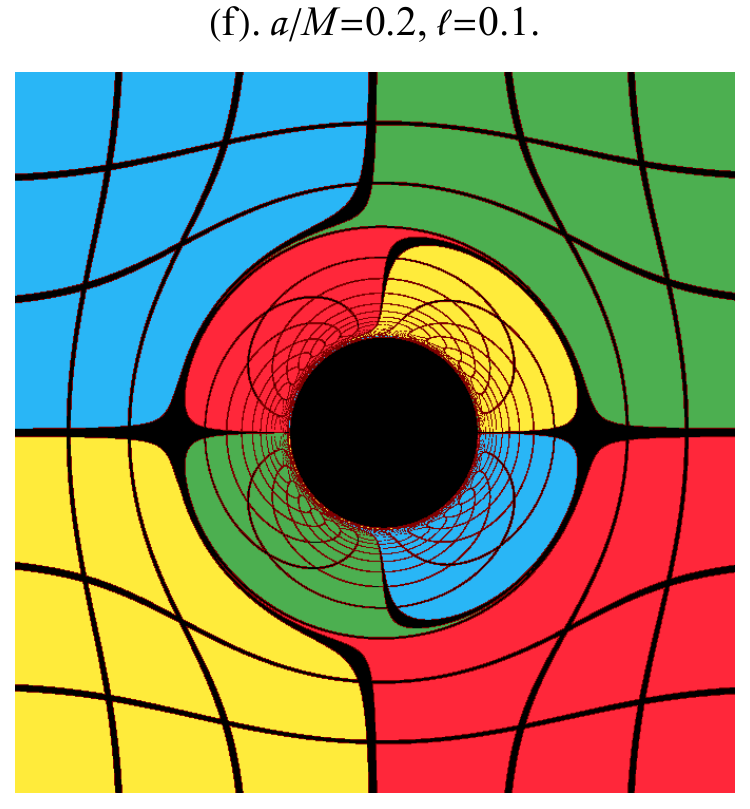}
\includegraphics[width=0.22\linewidth]{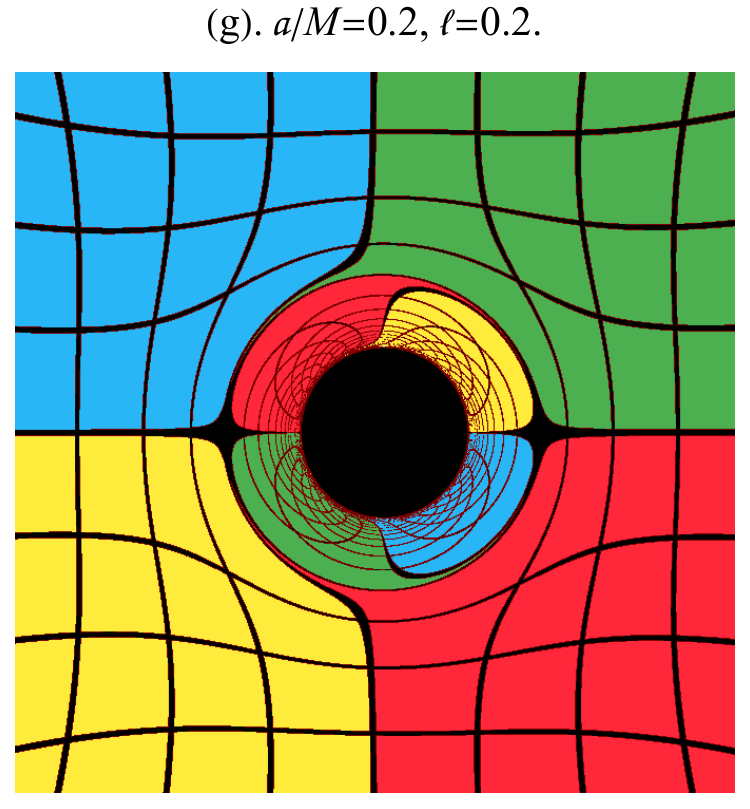}
\includegraphics[width=0.22\linewidth]{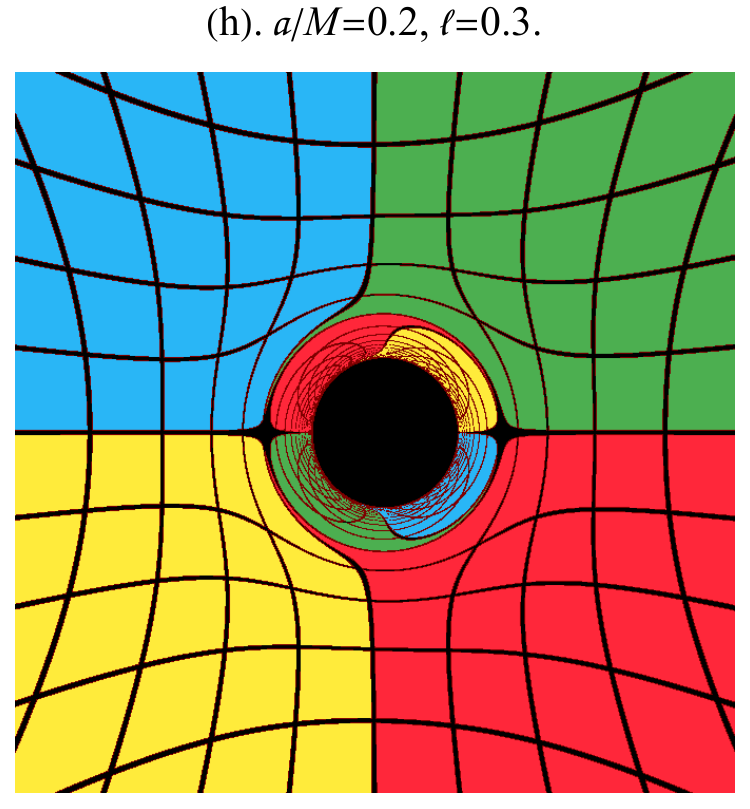}
\includegraphics[width=0.22\linewidth]{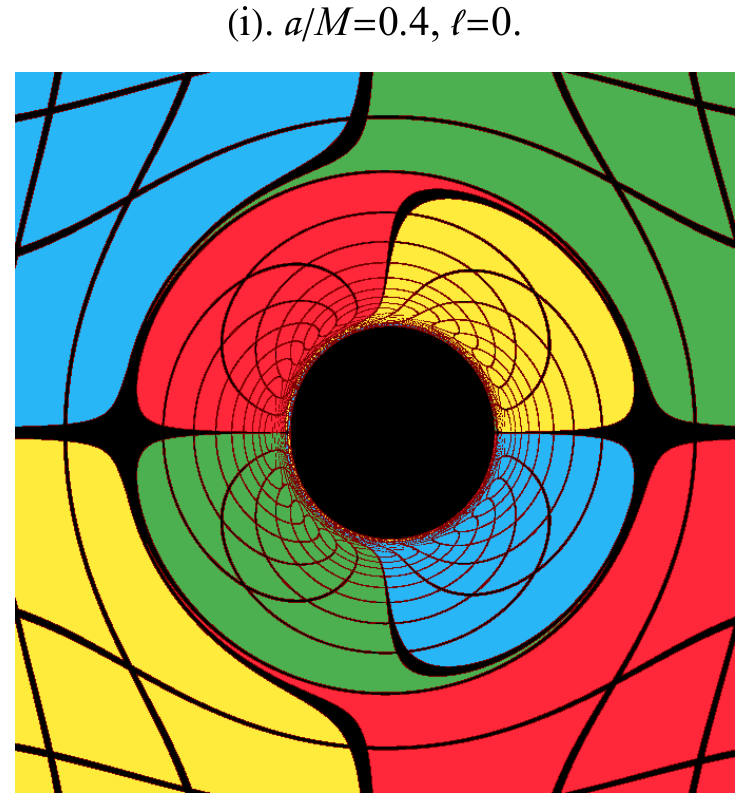}
\includegraphics[width=0.22\linewidth]{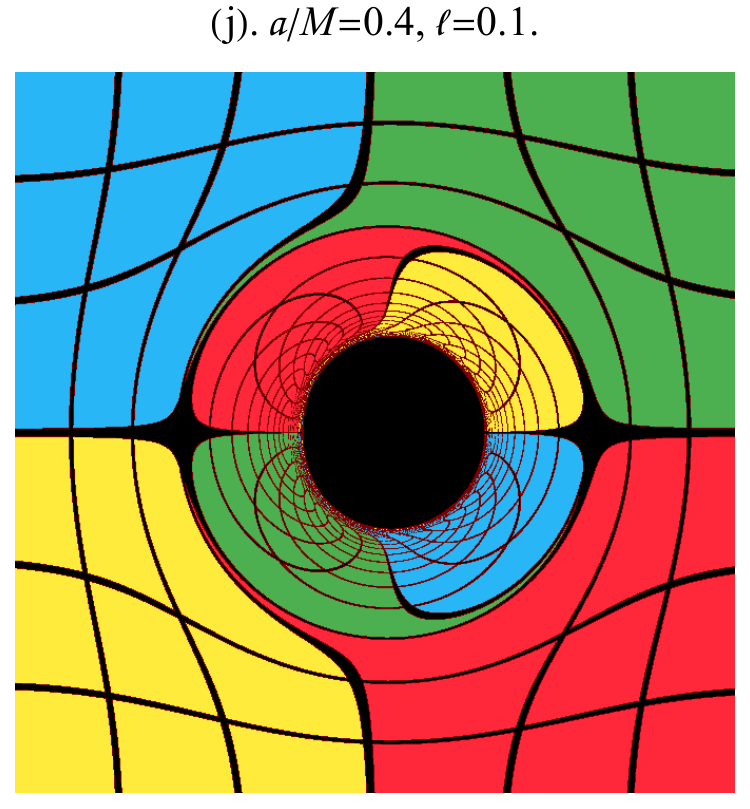}
\includegraphics[width=0.22\linewidth]{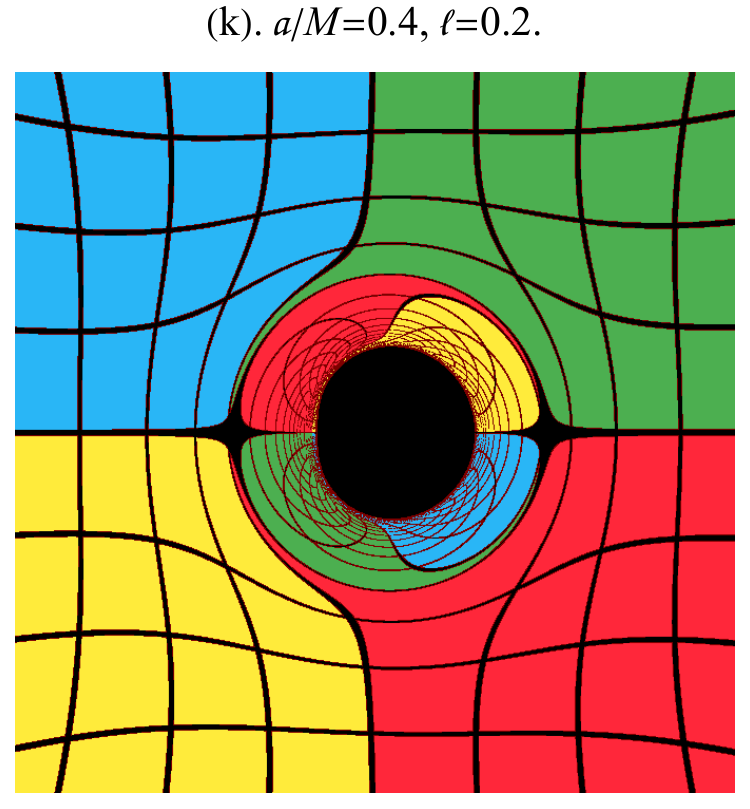}
\includegraphics[width=0.22\linewidth]{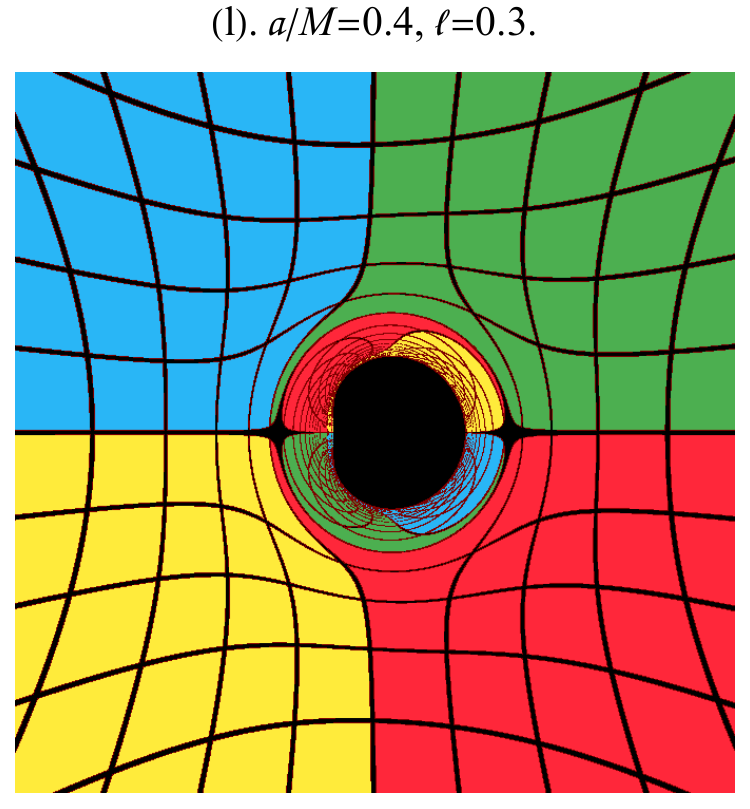}
\includegraphics[width=0.22\linewidth]{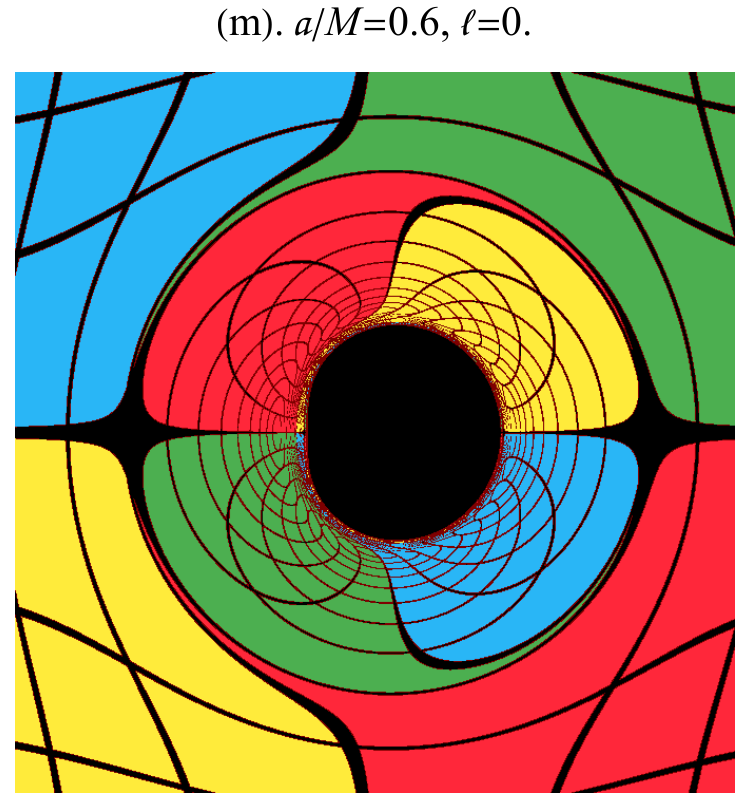}
\includegraphics[width=0.22\linewidth]{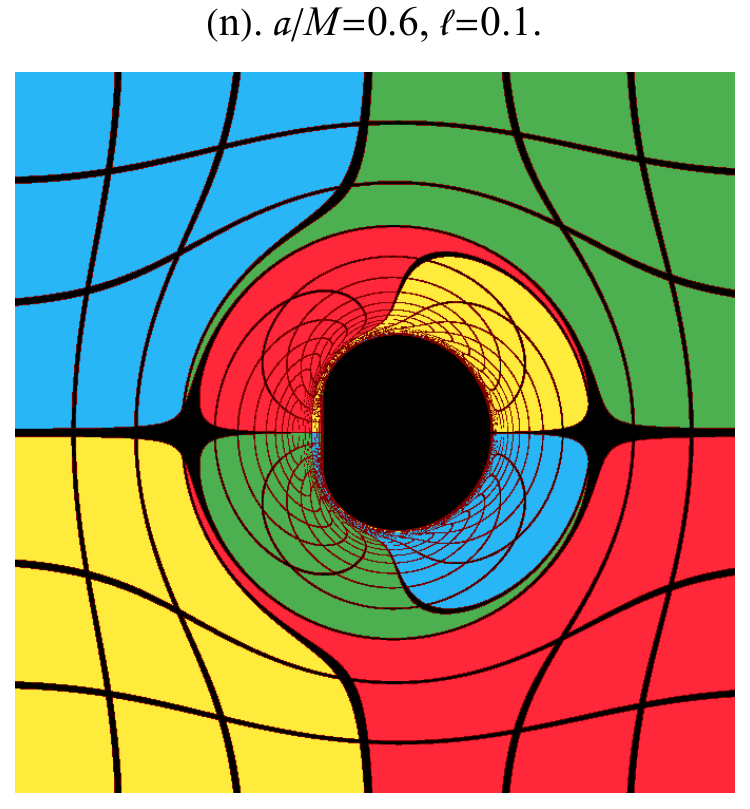}
\includegraphics[width=0.22\linewidth]{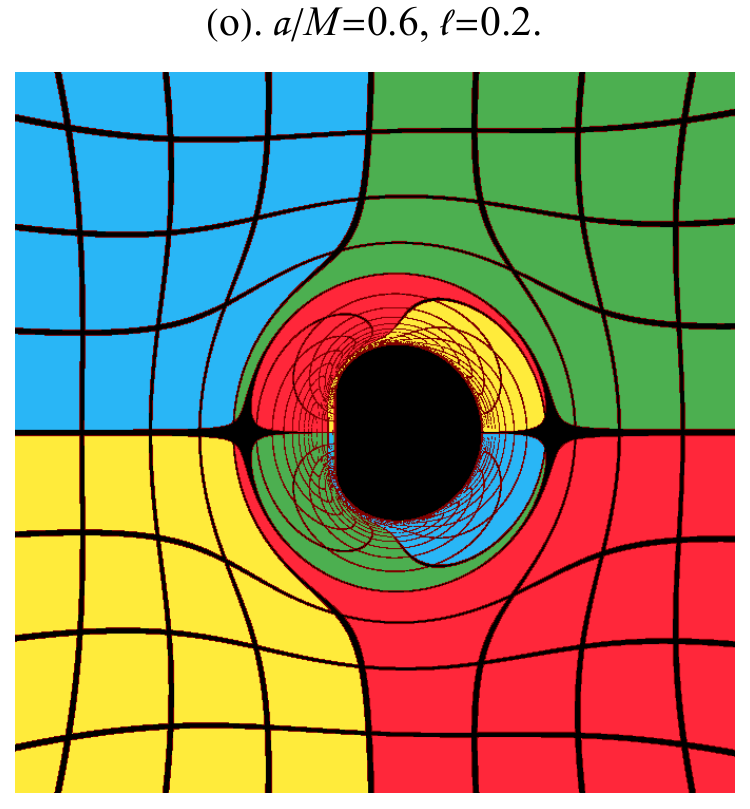}
\includegraphics[width=0.22\linewidth]{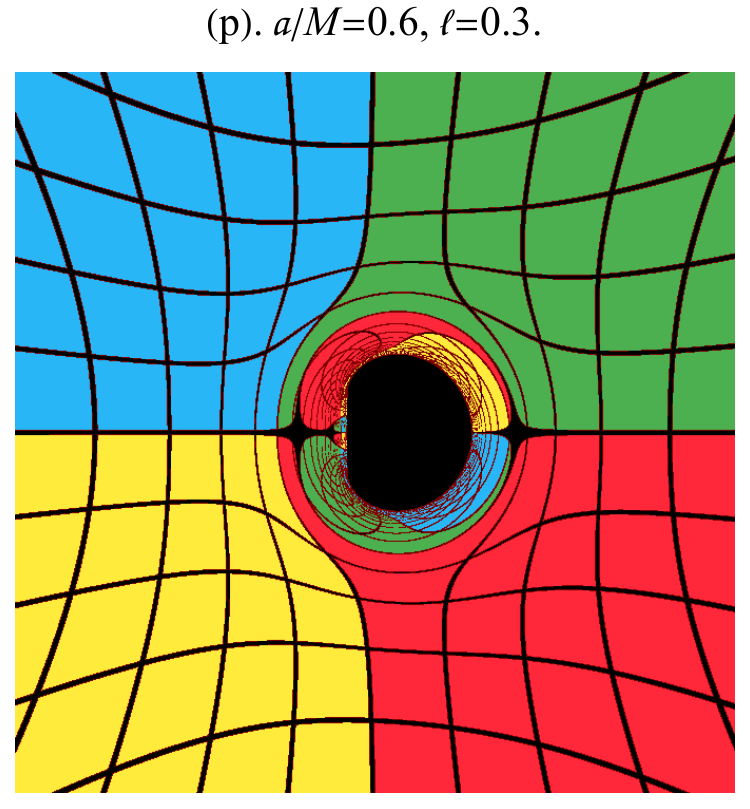}
\caption{Shadows cast by slowly rotating black holes with a Lorentz violation parameter $ \ell $ (Case B), as seen by an observer at $ \theta_0=\pi/2 $.}
\label{fig4}
\end{figure*}

Hence, we have discussed the impact of Lorentz violation paramter $\ell$ on the black hole shadow (Case B), within the parameter range of $ a/M\leq0.6 $.
Meanwhile, we also aim to explore whether Lorentz violation contributes to the frame-dragging effect due to the rotation of spacetime.
To achieve this, we employ the numerical backward ray-tracing method which sets the number of pixels to $ n=2048 $.
In Figs. \ref{fig4}(a)-\ref{fig4}(i), we can observe that as the Lorentz violation parameter increases, the size of the black hole shadow becomes smaller.
Meanwhile, the effects of rotation, including the D-shaped structure and the frame-dragging effect, are also amplified.
The coupling between the KR field and gravity can result in slowly rotating black holes exhibiting the apparent characteristics of extremal black holes in GR.
While the KR field seems to ``accelerate" the spin of the black hole, it also causes a significant gravitational lensing effect, with the shadow occupying a larger area within the photon ring, which can serve as a criterion to distinguish KR gravity from GR.

\subsection{Observable measurements}\label{Sec.4.2}

Thus far, we have established that the spin parameter $\bar{a}$ and the Lorentz violation parameter $\ell$ have a significant impact on the apparent shape of the shadow in slowly rotating KR spacetime.
To effectively characterize the shadow, it is crucial to use measurable and reliable properties.
We utilize two such observable measurements: the radius $ R_s $ and the distortion parameter $ \delta_s $, as outlined in Ref. \cite{Hioki:2009na}.
\begin{figure}[h]
\centering
\includegraphics[width=0.85\linewidth]{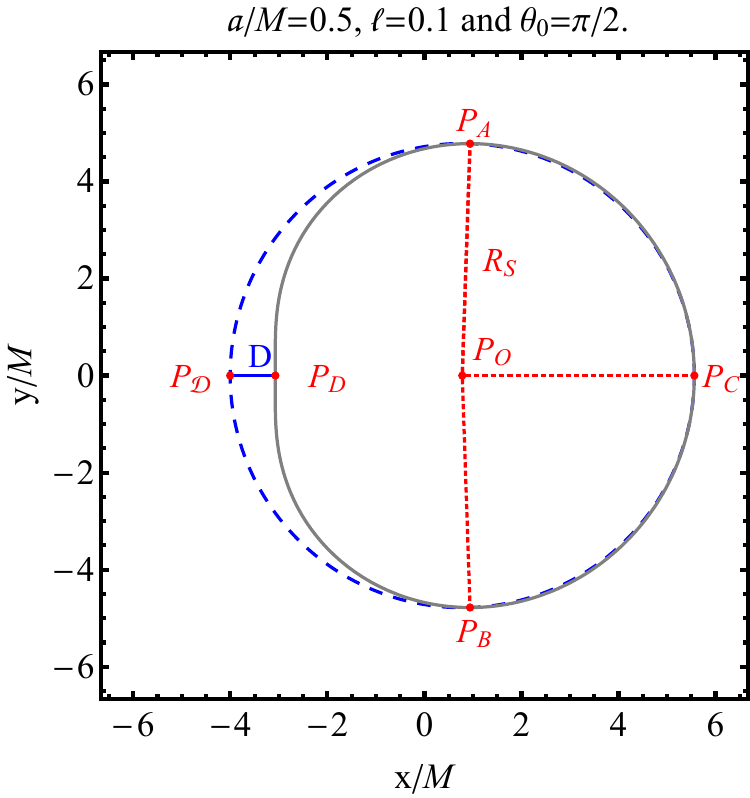}
\caption{ The observable measurements for the apparent shape of the black hole are the radius of the reference circle $ R_s $ and the distortion parameter $ \delta_s=\text{D}/R_s $, where $ \text{D} $ is the difference between the left endpoints of the reference circle and of the shadow.}
\label{fig5}
\end{figure}

The size of the shadow is characterized by the parameter $ R_s $, which corresponds to the radius of the reference circle shown as a blue dashed line in Fig. \ref{fig5}.
Passing through three points, the reference circle intersects the top position $P_A=(x_t,y_t)$, the bottom position $P_B=(x_b,y_b)$, and the point $P_C=(x_r, 0)$.
The point $P_C$ corresponds to the unstable retrograde circular orbit as seen from the equatorial plane by an observer.
Additionally, the point $ P_O=(x_o,0) $ represents the center of the reference circle, and $ x_o $ can be derived from the coordinates of $ P_A, P_B $, and $ P_C $:
\begin{align}
x_o=\frac{x^2_r-x^2_t-y^2_t}{2(x_r-x_t)}.
\end{align}
Next, the difference between the shaded left-hand point $ P_D=(x_d,0) $ and the reference circle's left-hand point $ P_\mathcal{D}=(\tilde{x}_r,0) $ needs to be considered.
The size of this difference is evaluated by $ \mathrm{D}=|x_d-\tilde{x}_r|$.
Further, the two observable measurements are defined as follows \cite{Tang:2022uwi}:
\begin{align}
R_s=&\frac{(x_t-x_r)^2+y_t^2}{2|x_r-x_t|}, \\ \delta_s=&\frac{\mathrm{D}}{R_s}.
\end{align}
\begin{figure}[h]
\centering
\includegraphics[width=0.48\linewidth]{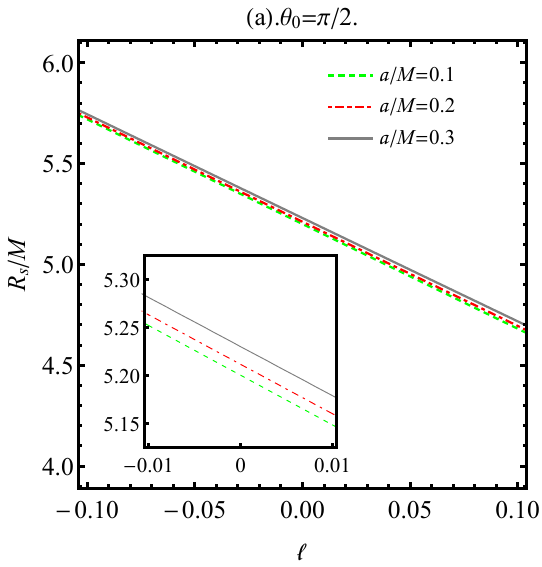}
\includegraphics[width=0.48\linewidth]{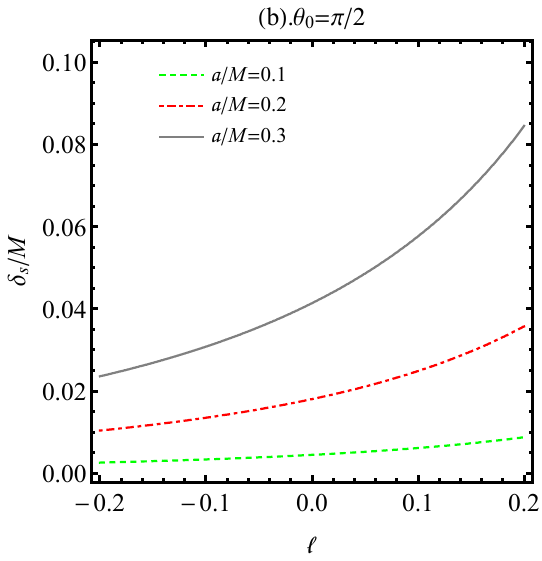}
\includegraphics[width=0.48\linewidth]{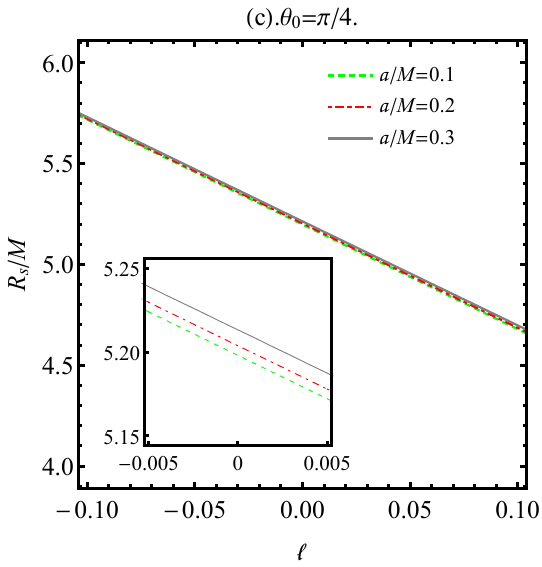}
\includegraphics[width=0.48\linewidth]{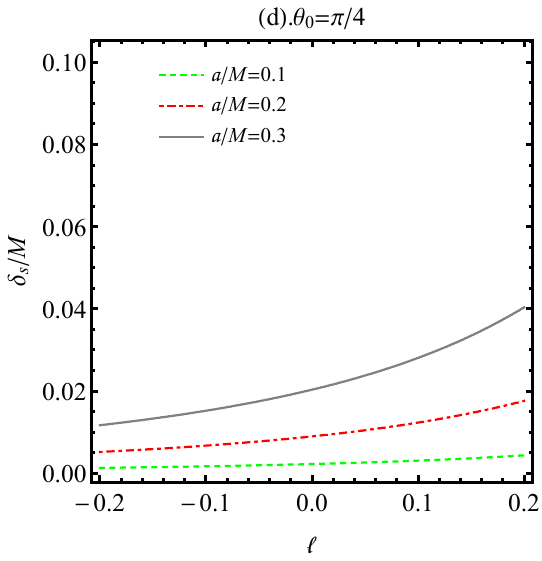}
\caption{Observable measurements $ R_s/M $ (left) and $ \delta_s/M $ (right) as functions of $ \ell $ are presented, respectively, for $ \theta_0=\pi/2 $ (top) and $ \theta_0=\pi/4 $ (bottom). }
\label{fig6}
\end{figure}

We numerically calculated these two observable measurements with the observation angle fixed at $ \theta_0=\pi/2 $ and $ \theta_0=\pi/4 $, respectively.
The results are presented in Figs. \ref{fig6}(a)-\ref{fig6}(d).
Note that the static case is not included in the results because the contour of the shadow coincides with the reference circle, resulting in no distortion.
Figs. \ref{fig6}-(b) and \ref{fig6}-(d) show that the observation angle $\theta_0$ has a significant effect on $\delta_s$.
Furthermore, $\delta_s$ increases as the parameter $\ell$ increases, suggesting that with a gravitationally coupled antisymmetric tensor field $B_{ab}$, a positive coupling constant $ (\xi_2>0) $ will cause the black hole shadow to become more deformed; conversely, a negative coupling constant $ (\xi_2<0) $ will decrease the deformation.
Figs. \ref{fig6}-(a) and \ref{fig6}-(c) clearly show the radius $R_s$ of the shadow.
It is evident from these results that when the black hole is rotating, a larger observation angle $ \theta_0 $ leads to a larger observed reference circle radius $ R_s $.
The effect of Lorentz violation results in a proportional correction to the radius $ R_s $, which decreases as $\ell$ increases.
This is consistent with the peculiar phenomena observed in Figs. \ref{fig4}(a)-\ref{fig4}(i) describing the shadow cast by slowly rotating KR black holes in Sec. \ref{Sec41}.
Interestingly, this trend is exactly the opposite of what is described in Eqs. \meq{XXX}-\meq{YYY}, because the effect of the KR field on the coordinates $ (X,Y) $ is smaller than the correction to the impact parameters.

For a approximatively estimation, utilizing slowly rotating KR metric \meq{caseB}, we calculate the angular radius of a black hole shadow, defined as $ \theta_{\text{BH}} = R_s \frac{\mathcal{M}}{D_O} $, with $D_O$ representing the distance from the observer to the black hole.
Specifically, for a black hole with mass $ \mathcal{M} $ located at a distance $ D_O $ from the observer, the angular radius $ \theta_\text{BH} $ can be expressed as  $ \theta_\text{BH} = 9.87098 \times 10^{-6} R_s \left(\frac{\mathcal{M}}{M_\odot}\right) \left(\frac{1 \text{kpc}}{D_O}\right) \mu\text{as} $ \cite{Amarilla:2011fx}.
\begin{table*}[ht]
\renewcommand{\arraystretch}{1.25}
\centering
\setlength\tabcolsep{1.3mm}{
\begin{tabular}{cccccccccccc}
\hline\hline
\multirow{2}{*}{$ \theta_\text{BH}(\mu arcsec) $} & \multirow{2}{*}{$\ell=-0.10$} & \multirow{2}{*}{$\ell=-0.08$}& \multirow{2}{*}{$\ell=-0.06$}& \multirow{2}{*}{$\ell=-0.04$}& \multirow{2}{*}{$\ell=-0.02$}& \multirow{2}{*}{$~~\ell=0~~$}& \multirow{2}{*}{$~\ell=0.01$}& \multirow{2}{*}{$~\ell=0.02$}& \multirow{2}{*}{$~\ell=0.03$}
\\ \\
\hline
  $a/M=0.0   $  &27.1905   & \textbf{26.6961}   & \textbf{26.2017}   & \textbf{25.7074}   & \textbf{25.2130}   &24.7186             &24.4714            &24.2242    &23.9771       \\
  $a/M=0.1   $  &27.2057   & \textbf{26.7116}   & \textbf{26.2179}   & \textbf{25.7243}   & \textbf{25.2307}   &24.7371             &24.4901            &24.2431    &23.9965         \\
  $a/M=0.2   $  &27.2505   & \textbf{26.7584}   & \textbf{26.2664}   & \textbf{25.7745}   & \textbf{25.2826}   & \textbf{24.7908}   &24.5457            &24.2998    &24.0540          \\
  $a/M=0.3   $  &27.3243   & \textbf{26.8342}   & \textbf{26.3451}   & \textbf{25.8562}   & \textbf{25.3673}   & \textbf{24.8785}   &24.6352            &24.3909    &24.1466            \\
  $a/M=0.4   $  &27.4237   & \textbf{26.9376}   & \textbf{26.4515}   & \textbf{25.9665}   & \textbf{25.4816}   & \textbf{24.9967}   & \textbf{24.7548}  &24.5129    &24.2709             \\
\hline\hline
\end{tabular}}
\caption{The numerical estimation of the angular radius of the supermassive black hole Sgr A* in our galaxy using the metric of a slowly rotating KR black hole. }
\label{tab1}
\end{table*}
\begin{table*}[t]
\renewcommand{\arraystretch}{1.25}
\centering
\setlength\tabcolsep{1.3mm}{
\begin{tabular}{cccccccccccc}
\hline\hline
\multirow{2}{*}{$ \theta_\text{BH}(\mu arcsec) $} & \multirow{2}{*}{$\ell=-0.15$} & \multirow{2}{*}{$\ell=-0.12$}& \multirow{2}{*}{$\ell=-0.09$}& \multirow{2}{*}{$\ell=-0.06$}& \multirow{2}{*}{$\ell=-0.03$}& \multirow{2}{*}{$~~\ell=0~~$}& \multirow{2}{*}{$~\ell=0.01$}& \multirow{2}{*}{$~\ell=0.02$}& \multirow{2}{*}{$~\ell=0.03$}
\\ \\
\hline
  $a/M=0.0   $  &22.8215   & \textbf{22.2262}   & \textbf{21.6308}   & \textbf{21.0355}   & \textbf{20.4401}   &\textbf{19.8448}   &\textbf{19.6463}   &19.4479          &19.2494        \\
  $a/M=0.1   $  &22.8326   & \textbf{22.2379}   & \textbf{21.6432}   & \textbf{21.0485}   & \textbf{20.4538}   &\textbf{19.8596}   &\textbf{19.6613}   &19.4630          &19.2651         \\
  $a/M=0.2   $  &22.8659   & \textbf{22.2726}   & \textbf{21.6799}   & \textbf{21.0874}   & \textbf{20.4950}   &\textbf{19.9027}   &\textbf{19.7059}   &\textbf{19.5086} &19.3112          \\
  $a/M=0.3   $  &22.9196   & \textbf{22.3295}   & \textbf{21.7396}   & \textbf{21.1506}   & \textbf{20.5618}   &\textbf{19.9732}   &\textbf{19.7778}   &\textbf{19.5817} &19.3855           \\
  $a/M=0.4   $  &22.9934   & \textbf{22.4069}   & \textbf{21.8214}   & \textbf{21.2360}   & \textbf{20.6515}   &\textbf{20.0680}   &\textbf{19.8738}   &\textbf{19.6796} &19.4854            \\
\hline\hline
\end{tabular}}
\caption{The numerical estimation of the angular radius of the supermassive black hole M87* using the metric of a slowly rotating KR black hole.}
\label{tab2}
\end{table*}
Tables \ref{tab1} and \ref{tab2} show the calculated angular radius of the black holes at the Galactic center (Sgr A*) and in the galaxy M87 (M87*), respectively, using the slowly rotating KR metric.
In this analysis, we use the latest observations indicating that the mass of the black hole Sgr A* is $\mathcal{M} = 4.0 \times 10^6 M_\odot$ with an observer distance of $D_O = 8.3$ kpc \cite{EventHorizonTelescope:2022wkp}.
For the black hole M87*, the mass is $\mathcal{M} = 6.5 \times 10^9 M_\odot$ with an observer distance of $D_O = 16.8$ Mpc \cite{EventHorizonTelescope:2019ggy}.
It is important to note that we have highlighted the data that fall within the observed range of angular diameters for the Sgr A* and M87* black holes based on recent observations.\footnote{The angular diameters of the Sgr A* and M87* black holes have been measured as $51.8 \pm 2.3 \mu\text{as}$ \cite{EventHorizonTelescope:2022wkp} and $42 \pm 3.0 \mu\text{as}$ \cite{EventHorizonTelescope:2019dse}, respectively.}
The inclusion of the black hole spin provides a larger parameter space for the Lorentz violation parameters.
For the case of a positive coupling constant $ (\xi_2>0) $, the latest observational data from Sgr A* imposes a stringent constraint on the parameter, with $ \ell \lesssim 0.01 $.
However, for the case of a negative coupling constant $ (\xi_2<0) $, the Lorentz violation parameter has a wider range, with $ \ell \gtrsim -0.08 $.

\section{CONCLUSIONS AND EXTENSIONS} \label{Sec.5}


The KR gravity is an important theory involving nonminimal coupling with an antisymmetric tensor field, resulting in spontaneous Lorentz symmetry breaking.
This theory has two types of static spherically symmetric vacuum solutions, with or without cosmological constants \cite{Liu:2024oas}.
We extend the solutions to include a spin $ a=J/M $ that describes the angular momentum of the black hole in both Case A and Case B.
Although the black hole metric is solved using the slow rotation approximation, we find that this approximation works very well for studying the black hole shadow contours when the black hole spin is in the range $ a/M \lesssim 0.4 $.
The validity of this approximation is also confirmed in real astronomical environments.
Insights from the analysis proposed by Renolds et al. of the LIGO-Virgo strain data of the 10 binary black hole mergers reveal that typical spins are constrained to $ a/M\lesssim0.4 $, even if the underlying population has randomly oriented spins \cite{Roulet:2018jbe}.
Therefore, our study is reliable in astronomical applications and can be used as an alternative to the shadow of exact axisymmetric KR black holes to test gravitational theories and constrain the upper limits on Lorentz violations.

For the shadow of the slowly rotating KR black holes, the photon trajectories in Case A are equivalent to those in a slowly rotating Kerr black hole, while the effects of Lorentz violation could exist in higher orders of rotation, i.e., $ \mathcal{O}(\tilde{a}^2) $.
In Case B, Lorentz violation leads to several novel consequences, summarized as follows: as the Lorentz violation parameter increases, the size of the black hole shadow decreases; the shape of the shadow becomes more bulging, causing the contours of black holes to approach those of extreme rotating black holes.
This change impacts the frame-dragging effect induced by rotation, resulting in greater distortion of space compared to that observed in GR. 
For observable measurements, we consider the radius $R_s$, which is associated with the apparent size, and the distortions $\delta_s$, which relate to the deformation of the shadow.
We find that the interplay between the black hole spin and the Lorentz violation parameter leads to an intriguing phenomenon in the black hole shadow contours: i.e., the Lorentz violation parameter $\ell$ proportionally alters the radius $R_s$, while the deformation $\delta_s$ exhibits an accelerated change.
Additionally, we used the latest observational data from M87* and Sgr A* to impose constraints on the Lorentz violation parameter $(-0.08 \lesssim \ell \lesssim 0.01)$ for KR black holes, thereby confirming the possibility of spontaneous Lorentz violation in spacetime.

There are several promising topics to be pursued in the future.
One intriguing direction is to investigate additional observable characteristics of the shadow cast by a rotating KR black hole.
To address potential shadow degeneracy caused by black hole parameters, we can examine other distortion parameters, such as the oblateness parameter $K_s$ and the shadow's "thickness" parameter $T$ \cite{Wang:2018eui}.
Additionally, to visualize the shadow in a realistic astronomical context, it is important to consider the emissions from the accretion disk surrounding the black hole \cite{Rosa:2023qcv,Rosa:2023hfm,Rosa:2024bqv,Huang:2024eir}.
Another promising area for future research is the exploration of topological classifications \cite{Wei:2022dzw,Wu:2022whe,Wu:2023sue,Wu:2023xpq,Wu:2023fcw,Wu:2023meo,Zhu:2024jhw} and phase transition criticality \cite{Hawking:1982dh,Kubiznak:2012wp,Ahmed:2023dnh,Wu:2024rmv} of the rotating black holes in KR gravity.
Finally, as is well known, when black holes have angular momentum, they exhibit additional properties such as superradiant instability \cite{Berti2009}, boson clouds \cite{Cannizzaro:2023jle,Lei:2023wlt}, and Proca clouds \cite{Pani2012prl}.
These phenomena are part of alternative theories to explain dark matter \cite{Branco:2023frw} and may potentially form so-called "black hole bombs" under specific conditions \cite{Feiteira:2024awb,Hawking:1974rv}.
Investigating the potential impact of Lorentz violation on the aforementioned phenomena using the slowly rotating solutions provided in this paper is also highly worthwhile, as such studies are expected to significantly advance our understanding of the intrinsic nature of gravity.

\acknowledgments

This work is supported by the National Natural Science Foundation of China (NSFC) under Grants No. 12205243, No. 12375053, No. 12122504, and No. 12035005; the Sichuan Science and Technology Program under Grant No. 2023NSFSC1347; the Doctoral Research Initiation Project of China West Normal University under Grant No. 21E028; the innovative research group of Hunan Province under Grant No. 2024JJ1006; and the Hunan Provincial Major Sci-Tech Program under grant No.2023ZJ1010.

\end{document}